\documentclass[a4paper,11pt]{article}

\usepackage{jcappub} 

\usepackage[T1]{fontenc} 

\usepackage{amsmath,amssymb,amsfonts}
\usepackage{mathtools}
\usepackage{bm}
\usepackage{bbm}

\usepackage{graphicx}
\usepackage{xcolor}
\usepackage{caption}

\usepackage{comment}

\usepackage{natbib}
\usepackage{aas_macros}

\usepackage{hyperref}
\hypersetup{unicode=true}

\usepackage{amsthm}
\usepackage{tikz}
\usetikzlibrary{arrows.meta,positioning,calc,fit,cd,decorations.pathmorphing,intersections}

\theoremstyle{definition}

\theoremstyle{plain}

\theoremstyle{plain}
\newtheorem{proposition}{Proposition}[section]

\theoremstyle{plain}

\theoremstyle{remark}

\theoremstyle{remark}

\theoremstyle{remark}

\def\pa#1#2{\dfrac{\partial #1}{\partial #2}}

\def\pd{{\rm d}}
\def\Tr{{\rm Tr}}

\def\deltadir{\delta_{\rm D}}

\title{Curvature-Conditioned Measures for Cosmological Peak Statistics:\\
A Transport-Geometric Framework}

\author[a,b,1]{Tsutomu T. Takeuchi,\note{Corresponding author.}}

\affiliation[a]{Division of Particle and Astrophysical Science, Nagoya University, Furo-cho, Chikusa-ku, Nagoya, Aichi 464--8602, Japan}
\affiliation[b]{The Research Center for Statistical Machine
Learning, the Institute of Statistical Mathematics, 10--3 Midori-cho, Tachikawa, Tokyo 190--8562, Japan}

\emailAdd{tsutomu.takeuchi.ttt@gmail.com}

\abstract{
We develop a transport-geometric theory of cosmological peak statistics based on optimal transport and entropy geometry.
The density field is treated as a probability measure in Wasserstein space, and its local structure is characterized by a logarithmic curvature tensor obtained as the localized response of an entropy functional.
Peaks are thereby defined as positive-curvature stationary points, and their abundance is formulated as a curvature-conditioned measure on local tensor space.
In the Gaussian-linear limit, this measure admits a finite-dimensional closure in terms of the spectral moments of the density field.
The resulting peak abundance reduces exactly to the classical BBKS formula, identifying BBKS as a solvable Gaussian closure of a more general geometric structure.
This formulation separates peak statistics into three fundamental ingredients: the probability distribution of local variables, the positive-curvature constraint, and the induced geometric measure.
The theory extends naturally beyond the Gaussian approximation.
Nonlinear evolution appears as a deformation of the logarithmic curvature geometry, while primordial non-Gaussianity is interpreted as a deformation of the curvature-conditioned measure itself.
We further formulate two- and three-point peak statistics as higher-order curvature-conditioned measures and show that the resulting hierarchy can be organized as response functions to long-wavelength background modes, with conventional peak bias emerging as the lowest-order response coefficient.
These results provide a unified description linking optimal transport, curvature geometry, peak statistics, and cosmological observables, and establish a systematic framework for studying nonlinearity, scale dependence, primordial non-Gaussianity, and higher-order peak correlations.
}

\begin{document}

\maketitle

\flushbottom

\section{Introduction}
\label{sec:introduction}

The large-scale structure of the Universe is characterized by the growth of primordial density fluctuations and their statistical properties.
Among these, local maxima of the density field, namely peaks, are considered to play an important role as seeds of dark matter halos and galaxy formation, and their statistical properties have been studied for many years.
In this context, \cite{1986ApJ...304...15B} (hereafter BBKS) is widely known as a classical theory that systematically provides peak statistics in a Gaussian random field.
BBKS derived, in closed form, the peak number density, the curvature distribution, and the peak bias based on the gradient and the Hessian of the density fluctuation field, and has since served as a fundamental reference point for peak theory \citep{1986ApJ...304...15B}.
Furthermore, the relation between peak bias and halo formation has been extended in connection with the ideas of peak-background split and excursion set, and has become one of the foundations of structure formation theory \citep[e.g.,][]{1996MNRAS.282..347M, 1996MNRAS.282.1096M, 1999MNRAS.308..119S, 2010PhRvD..81b3526D}.

On the other hand, in recent theories of large-scale structure, the description of bias itself has undergone significant development.
In particular, integrated perturbation theory (iPT) based on Lagrangian perturbation theory provides a framework that treats arbitrary bias models in a unified manner \citep{2011PhRvD..83h3518M, 2012PhRvD..86f3518M}, and more recent studies have clarified that bias can be formulated not only as a function of density but also as a function of local tensor fields including its gradients and second derivatives \citep{2019PhRvD.100h3504M, 2024PhRvD.110f3543M}.
That is, the distribution of galaxies and halos can be expressed as a semi-local bias of the form
\begin{align}
	F = F(\delta,\partial_i\delta,\partial_i\partial_j\Phi,\dots)
\end{align}
and it has been shown that tidal tensors and shape information are essentially involved in the statistics.
Such developments indicate that cosmological statistics are being extended from scalar quantities to tensor quantities.

Nevertheless, even through developments since BBKS, the theoretical starting point of peak statistics still strongly depends on the assumptions of Gaussianity, linear theory, and smoothing at a fixed scale.
Under these assumptions, the joint distribution of local variables closes, and calculations can be carried out in a clear form.
On the other hand, it is not easy to treat nonlinear structures, non-Gaussian initial conditions, scale dependence, and $n$-point statistics in a unified manner.
Therefore, it is required to reconsider peak statistics at a level prior to these assumptions and to clarify which structures are essential and which parts are specific to the Gaussian limit.

This line of thought is different from simply adding correction terms to the BBKS theory.
Rather, what is required is to reposition peaks not as a given collection of density maxima, but within a more general geometric or measure-theoretic structure.
In fact, cosmological density fields are given observationally as a finite number of galaxies or halos, and their statistical description intrinsically involves the problem of finite samples and finite observational windows \citep{takeuchi2026finite_sample_window}.
Moreover, from a theoretical point of view, structure formation is not merely a change in field values, but has an aspect that should be understood as the spatial rearrangement of mass distributions.
This suggests that peak statistics can be regarded not only as an extremum problem in functional analysis, but also as a problem of measure deformation and local geometry.

Furthermore, it is important that such a reformulation is carried out on the same space of local variables as recent tensor bias theories.
That is, the set of local quantities $(\delta,\partial_i\delta,\partial_i\partial_j\delta)$ serves as the fundamental variables characterizing bias in iPT and its extensions \citep{2024PhRvD.110f3543M}.
In these approaches, these quantities are typically treated as arguments of a bias function, whereas in the present work they are regarded as random variables, with a measure structure introduced on this space.
This distinction leads to complementary descriptions on the same local tensor space, namely ``bias as a function'' and ``peak statistics as a measure.''

From this perspective, optimal transport theory naturally emerges.
Optimal transport is a framework that measures the difference between two distributions not as a difference of function values but as a cost of mass rearrangement \citep{monge1781memoire, kantorovich1942translocation, villani2003, villani2009optimal}, and provides a natural language to describe cosmological structure formation as a deformation of mass distributions.
In particular, once one enters the space of probability measures, the deformation of distributions is described geometrically by the Wasserstein distance, and the behavior of entropy functionals plays an important role \citep{Otto2001, lott2009ricci, sturm2006geometryI, sturm2006geometryII}.
Although this structure has not been emphasized in conventional peak theory, it provides a powerful clue to embed the statistics of local extrema into a more general geometry.

Furthermore, in recent cosmology, the viewpoint of understanding structure formation itself as transport or deformation is becoming increasingly important.
For example, Lagrangian mass transport, the geometry of displacement fields, and statistical measures based on distances between distributions characterize structure from perspectives different from conventional correlation functions and power spectra \citep{2003MNRAS.346..501B, Takeuchi2026OptimalTransport}.
Within this trend, reconsidering peak statistics in connection with optimal transport is not merely a formal generalization, but has significance as an attempt to reorganize the foundations of structure formation theory.
To our knowledge, however, the classical theory of peak statistics, known as BBKS, has not been reformulated directly within the framework of optimal transport, Wasserstein geometry, or entropy-based curvature.
Existing developments of peak theory have primarily focused on extensions within the random-field and bias-based frameworks, while applications of optimal transport in cosmology have been directed mainly toward reconstruction and mass transport problems.
In this sense, a direct connection between peak statistics and transport geometry has remained largely unexplored.

The starting point of this study is to regard the density field as a probability measure and to characterize its local structure from the standpoint of entropy geometry.
From this viewpoint, the Hessian used to describe local extrema is reinterpreted not as a merely analytical auxiliary quantity, but as a local response to deformations of the distribution.
As a result, questions such as what a peak is, why the determinant of the Hessian appears, and why BBKS-type formulas close in the Gaussian limit are understood not as individual computational techniques but as consequences of a more general curvature statistics.
The aim of this paper is to make this perspective explicit and to reformulate peak statistics as a curvature-conditioned point process.

\begin{figure*}[t]
	\centering
	\includegraphics[width=\textwidth]{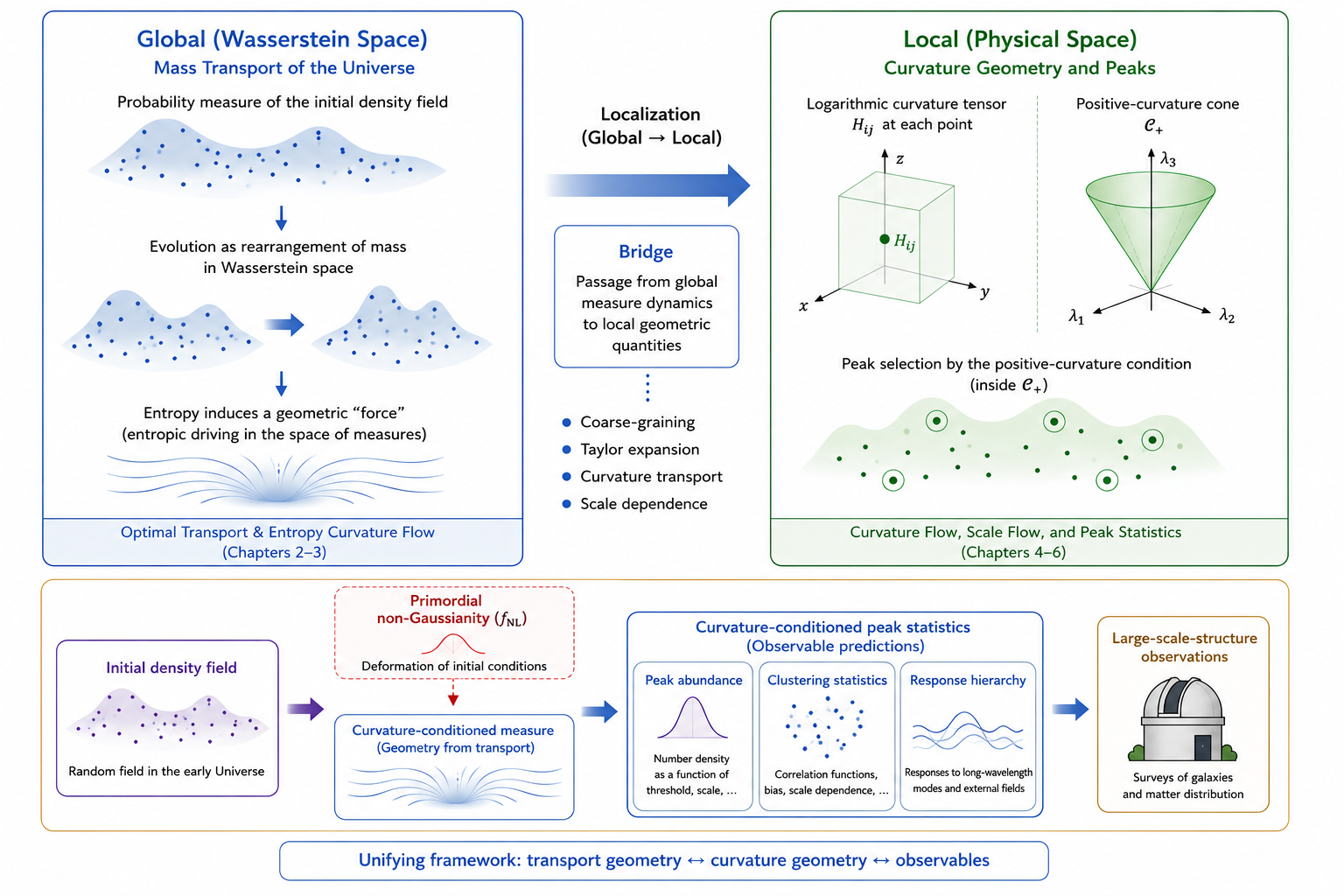}
	\caption{
	Conceptual overview of the transport-geometric framework developed in this work.
	The left panel illustrates cosmological structure formation as a rearrangement of mass distributions in Wasserstein space.
	In this picture, entropy-response geometry generates a global notion of curvature on the space of probability measures.
	Through localization, this global measure geometry gives rise to the logarithmic curvature tensor and the positive-curvature cone shown in the right panel.
	Peaks are then formulated as a curvature-conditioned point process selected by local positivity of the logarithmic curvature.
	The lower panel summarizes the connection between initial conditions and observables.
	Primordial non-Gaussianity acts as a deformation of the initial curvature-conditioned measure, while peak abundance, clustering statistics, and response hierarchies emerge as observable consequences of the same underlying geometric structure.
	This framework therefore provides a unified bridge between optimal transport, curvature geometry, peak statistics, and cosmological observations.
	}
	\label{fig:intro_overview}
\end{figure*}

The framework developed in this paper provides a geometric connection between mass transport, local curvature structure, and observable peak statistics.
Within this picture, the BBKS theory appears as the Gaussian-linear limit of a more general curvature-conditioned measure.
This perspective clarifies which aspects of peak statistics arise from Gaussian closure and which originate from the underlying geometric structure.
It also provides a natural setting in which nonlinear evolution,
primordial non-Gaussianity, scale dependence, and higher-order peak correlations can be discussed within a common geometric language. 
Rather than modifying the BBKS theory itself, the present approach aims to identify the broader geometric structure in which the familiar Gaussian results emerge as a tractable special case.

The conceptual structure of the framework is summarized in
Figure~\ref{fig:intro_overview}.
The paper then develops each component of this picture in turn.
In Sec.~\ref{sec:curvature_geometry}, we introduce the measure-theoretic description of the density field and derive the logarithmic curvature tensor from entropy-response geometry in Wasserstein space.
In Sec.~\ref{sec:peaks}, we construct the curvature-conditioned point process and show how the Gaussian closure reproduces the classical BBKS results.
In Secs.~\ref{sec:nonlinear_deformation} and \ref{sec:response_hierarchy}, we investigate nonlinear deformations, scale-dependent curvature flow, and extensions beyond the Gaussian limit while preserving the underlying geometric structure.
In Sec.~\ref{sec:beyond_gaussian}, we formulate peak correlations and response hierarchies as higher-order statistics of the curvature-conditioned point process, and discuss their connection to primordial non-Gaussianity.
Finally, in Sec.~\ref{sec:discussion}, we summarize the main results and outline future theoretical and observational directions.
Technical details of the eigenvalue-space representation of the curvature measure and its reduction to the BBKS formula are collected in Appendix~\ref{app:eigenvalue_measure}.

\section{Curvature Geometry of Cosmological Density Fields}
\label{sec:curvature_geometry}

\subsection{Measure-Theoretic Description of the Density Field}

The cosmological density field is usually described as a function $\rho(\bm{x})$ on space \citep[e.g.,][]{1980lssu.book.....P, Peacock1999Cosmology, takeuchi2025physics}, but in this work we reinterpret it as a probability measure \citep[cf.][]{Takeuchi2026OptimalTransport}.
In what follows, the spatial position vector is written as $\bm{x}\in\mathbb{R}^3$, while in measure-theoretic notation we use $\pd^3 x$ for brevity.
That is, on a finite domain $V\subset\mathbb{R}^3$, we define
\begin{align}
	\mu(\pd^3 x)=\rho(\bm{x})\pd^3 x.
\end{align}
If necessary, by imposing the normalization
\begin{align}
	\int_V \rho(\bm{x})\pd^3 x=1,
\end{align}
$\mu$ can be regarded as a probability measure.

The significance of this reformulation lies in viewing structure formation not as a ``change in field values,'' but as a ``rearrangement of mass distributions.''
In fact, galaxy distributions and dark matter distributions are observed as a finite number of particles or halos, and are therefore essentially given as point processes or finite measures \citep{takeuchi2026finite_sample_window}.
Thus, in order to describe the comparison and evolution of cosmological structures, it is natural to deal with differences between measures rather than differences of function values.
From this perspective, structure formation is described as the time evolution of a measure
\begin{align}
	\mu_t(\pd^3 x)=\rho(t,\bm{x})\pd^3 x,
\end{align}
and the problem reduces to how to define a distance between two measures.

This measure-theoretic viewpoint is also useful for clarifying what is meant by a local structure.
A peak of a density field is usually introduced through derivatives of a function, but the same object can also be viewed as a locally selected structure of a measure.
In this work, the derivatives that define peaks are not treated as purely analytic auxiliaries.
They will be derived from the local response of an entropy functional on the space of probability measures.
This distinction is important because it separates the kinematic condition for locating stationary points from the geometric origin of the curvature tensor used to characterize them.

\subsection{Optimal Transport and Distance Between Distributions}

The simplest distance between measures is the $L^p$ distance based on pointwise differences, but this does not capture the viewpoint of mass rearrangement.
For example, a spatial translation of a mass distribution produces a large $L^p$ distance even though the underlying structure is physically unchanged.

A natural framework that resolves this issue is optimal transport.
In optimal transport, the distance between two distributions is defined by the minimal cost required to transport one measure into the other.
In Monge's formulation, one considers a map $T$ satisfying
\begin{align}
	\nu=T_\#\mu,
\end{align}
and minimizes
\begin{align}
	\int |\bm{x}-T(\bm{x})|^2 \mu(\pd^3 x)
\end{align}
\citep{monge1781memoire}.
Here $T_\#\mu$ denotes the pushforward measure induced by $T$, defined by
\begin{align}
	T_\#\mu(A)=\mu(T^{-1}(A))
\end{align}
for any measurable set $A$.
Since the existence of an optimal map is not guaranteed in general, one instead uses the Kantorovich relaxation \citep{kantorovich1942translocation}.
Introducing a coupling measure $\pi\in\Pi(\mu,\nu)$, one defines
\begin{align}
	W_2^2(\mu,\nu)
	=
	\inf_{\pi\in\Pi(\mu,\nu)}
	\int |\bm{x}-\bm{y}|^2 \pi(\pd^3 x\,\pd^3 y).
\end{align}
Here $\Pi(\mu,\nu)$ denotes the set of coupling measures whose marginals are $\mu$ and $\nu$.
The resulting quantity $W_2$ defines a metric on the measure space $\mathcal{P}_2(\mathbb{R}^3)$ \citep{villani2009optimal}.

For the present purpose, however, optimal transport is important not merely as a distance between two density distributions.
More fundamentally, it equips the space of measures with a differential geometric structure.
In this geometry, infinitesimal deformations of a density are represented by transport velocities, and variations of functionals on measure space define gradient and Hessian structures.
The logarithmic curvature tensor introduced below will arise precisely from this variational framework.

\subsection{Wasserstein Geometry and Geodesics}

With the introduction of the distance $W_2$, the space of measures $\mathcal{P}_2(\mathbb{R}^3)$ acquires a geometric structure analogous to that of a Riemannian manifold \citep{villani2003,Otto2001}.
This space, called the Wasserstein space, is understood as a probability measure space endowed with the metric $W_2$ \citep{villani2009optimal}.
Within this framework, the shortest path connecting two distributions is defined as a geodesic.

Such geodesics are realized as displacement interpolations generated by a velocity field $\bm{v}(t,\bm{x})$ transporting the density $\rho(t,\bm{x})$.
The evolution obeys the continuity equation
\begin{align}
	\partial_t\rho+\nabla\cdot(\rho\bm{v})=0
\end{align}
\citep{BenamouBrenier2000},
which expresses mass conservation during transport.

A key property of Wasserstein geometry is that the transport velocity is not arbitrary.
Under suitable regularity conditions, it can be represented as the gradient of a scalar potential $\phi$,
\begin{align}
	\bm{v}=\nabla\phi
\end{align}
\citep{villani2009optimal,Otto2001}.
Consequently,
\begin{align}
	\partial_t\rho+\nabla\cdot(\rho\nabla\phi)=0
\end{align}
holds, showing that measure deformations are described by potential flows.
Structure formation is therefore interpreted geometrically as transport of mass distributions generated by a potential.

For later use, we also recall the infinitesimal form of this geometry.
A tangent vector at a smooth density $\rho$ can be represented by a potential $\phi$ through
\begin{align}
	\dot{\rho}=-\nabla\cdot(\rho\nabla\phi).
\end{align}
Thus, an infinitesimal displacement in Wasserstein space is not an arbitrary perturbation of $\rho$, but a mass-conserving rearrangement generated by a velocity potential.
This variational structure will play a central role below, where the logarithmic curvature tensor is derived from the response of entropy under such displacement perturbations.

\subsection{Entropy and Curvature of Measure Space}

The geometric structure in Wasserstein space is characterized by variations of functionals on measures.
In particular, a fundamental role is played by the entropy functional\footnote{To be precise, this is Boltzmann's $H$-function, a negative of usual entropy \citep{villani2009optimal}. 
}
\begin{align}
	\mathcal{E}(\rho)=\int \rho(\bm{x})\ln\rho(\bm{x})\,\pd^3 x
\end{align}
\citep{villani2003,Otto2001}.
This functional quantifies the spread of a distribution, and by considering its variation along geodesics, one can describe geometric properties of the deformation of distributions.
In particular, the convexity of $\mathcal{E}$ along Wasserstein geodesics corresponds to the curvature of the measure space \citep{lott2009ricci,sturm2006geometryI,sturm2006geometryII}.
That is, curvature is defined as the response of entropy to deformations of the distribution.

In this sense, curvature appears not as a local tensor on a Riemannian manifold, but first as the rate of change of a functional on the space of measures.
Therefore, in order to construct the curvature geometry of cosmological density fields, it is necessary to translate this global concept into local quantities of the density field.
This point connects with recent perturbative frameworks in which local tensor fields are used as arguments of bias functions \citep{2024PhRvD.110f3543M}, but in the present work the curvature tensor is introduced not from a functional expansion but as a local limit of entropy response.
Thus, the second-order tensor appearing here is not a variable of bias expansion, but a local quantity derived from the geometry of measure space.
Details of the formulations and related properties are extensively discussed in a companion paper \citep{takeuchi2026entropic_curvature_cosmo}.

We now make this statement explicit.
Let $\rho_s$ be a smooth curve in Wasserstein space generated by a velocity potential $\phi_s$:
\begin{align}
	\partial_s\rho_s+\nabla\cdot(\rho_s\nabla\phi_s)=0.
\end{align}
At $s=0$, we write $\rho_0=\rho$ and $\phi_0=\phi$.
The first variation of the entropy along this displacement is
\begin{align}
	\left.\frac{\pd}{\pd s}\mathcal{E}(\rho_s)\right|_{s=0}
	&=
	\int \left(1+\ln\rho\right)\left.\partial_s\rho_s\right|_{s=0}\,\pd^3 x
	\nonumber\\
	&=
	-\int \left(1+\ln\rho\right)\nabla\cdot(\rho\nabla\phi)\,\pd^3 x
	\nonumber\\
	&=
	\int \rho\,\nabla\ln\rho\cdot\nabla\phi\,\pd^3 x,
\end{align}
where boundary terms have been omitted, assuming either periodic boundary conditions or sufficiently rapid decay at the boundary.
This expression shows that the Wasserstein gradient of the entropy is represented by the potential $\ln\rho$.
Equivalently, the associated entropy force field is
\begin{align}
	F_i(\bm{x})=-\partial_i\ln\rho(\bm{x}).
\end{align}
The local curvature relevant for peak statistics is the spatial response of this entropy force.
Taking a derivative of $F_i$, we obtain
\begin{align}
	\partial_jF_i(\bm{x})
	=
	-\partial_j\partial_i\ln\rho(\bm{x}).
\end{align}
This motivates the definition of the logarithmic curvature tensor
\begin{align}
	H_{ij}(\bm{x})
	\equiv
	-\partial_i\partial_j\ln\rho(\bm{x}).
\end{align}
Thus, $H_{ij}$ is not introduced as an ad hoc replacement of the density Hessian.
It is the local response tensor of the entropy force field induced by displacement variations in Wasserstein space.
The same point can be summarized as follows.

\begin{proposition}[Variational origin of the logarithmic curvature tensor]
Let $\rho$ be a smooth positive density on a domain $V\subset\mathbb{R}^3$, and let
\begin{align}
	\mathcal{E}(\rho)=\int_V \rho\ln\rho\,\pd^3 x
\end{align}
be the entropy functional on Wasserstein space.
For a displacement perturbation generated by a potential $\phi$, the first variation of $\mathcal{E}$ is represented by the Wasserstein gradient potential $\ln\rho$.
The local linear response of the associated entropy force
\begin{align}
	F_i=-\partial_i\ln\rho
\end{align}
to an infinitesimal spatial displacement is
\begin{align}
	\partial_jF_i=H_{ij},
	\qquad
	H_{ij}=-\partial_i\partial_j\ln\rho .
\end{align}
Therefore, the logarithmic curvature tensor $H_{ij}$ is the local tensorial response associated with the Wasserstein entropy gradient.
\end{proposition}
\medskip
\noindent
This proposition gives the variational origin of the curvature tensor used throughout this paper.
The construction starts from a functional on the space of probability measures, takes its Wasserstein first variation, and then localizes the response of the resulting entropy force.
Consequently, the tensor $H_{ij}$ is not merely a convenient nonlinear transformation of the usual Hessian.
It is fixed by the entropy geometry of mass rearrangements.

\subsection{Curvature Tensor as a Local Limit}

The discussion so far has been formulated in the geometry of probability measures.
While this global description provides the natural setting for transport and entropy, observable peak statistics are ultimately defined through local quantities in physical space.
The conceptual relation between these two levels of description is illustrated in Figure~\ref{fig:intro_overview}.
The localization step translates the global entropy-response geometry into local tensor variables that can be evaluated at each spatial position.
In this sense, the logarithmic curvature tensor introduced below is the physical-space manifestation of the underlying Wasserstein geometry.
We now derive this local curvature representation explicitly.

Since the entropy-based curvature described above is essentially a global quantity, it is necessary to consider its local limit in order to describe local structures.
For this purpose, we expand the density field $\rho(\bm{x})$ around a point $\bm{x}_0$ and consider a local quadratic approximation.

The Taylor expansion of the logarithmic density is written as
\begin{align}
	\ln\rho(\bm{x})=\ln\rho(\bm{x}_0)+(\bm{x}-\bm{x}_0)_i\partial_i\ln\rho(\bm{x}_0)+\frac{1}{2}(\bm{x}-\bm{x}_0)_i(\bm{x}-\bm{x}_0)_j\partial_i\partial_j\ln\rho(\bm{x}_0)+\cdots .
\end{align}
In this expansion, the first derivative represents the local slope of the entropy gradient potential, and the second derivative represents its local response.
Therefore, consistently with the variational derivation above, we define the curvature tensor as the Hessian of the logarithmic density,
\begin{align}
	H_{ij}(\bm{x})=-\partial_i\partial_j\ln\rho(\bm{x}).
\end{align}
The negative sign is taken so that $H$ becomes positive definite at a local maximum.

With this definition, curvature is introduced not as the second derivative of the density field itself, but as a local response to deformations of the distribution.
Thus, this tensor is obtained as a local representation of entropy curvature in Wasserstein geometry.
Here, the fact that $\ln\rho$ naturally appears in the local limit of entropy geometry will become important later when discussing nonlinearity.
Moreover, in recent tensor bias theories, decomposition of local second-order tensors into rotationally invariant quantities plays a central role \citep{2024PhRvD.110f3543M}, whereas the distinctive feature of this work is to assign a curvature-geometric meaning to the space of local tensors.

The logarithmic form also immediately clarifies how the present curvature differs from the conventional density Hessian beyond the linear regime.
Using
\begin{align}
	\partial_i\partial_j\ln\rho
	=
	\frac{1}{\rho}\partial_i\partial_j\rho
	-
	\frac{1}{\rho^2}\partial_i\rho\,\partial_j\rho,
\end{align}
we obtain
\begin{align}
	H_{ij}
	=
	-\frac{1}{\rho}\partial_i\partial_j\rho
	+
	\frac{1}{\rho^2}\partial_i\rho\,\partial_j\rho .
\end{align}
The first term is the density Hessian weighted by the local density, whereas the second term is a positive semi-definite gradient-square contribution.
Therefore, in nonlinear regions where density gradients are large, the logarithmic curvature tensor contains information that is absent from the ordinary Hessian of $\rho$ or of the linear density contrast.
This term is not an optional correction; it is fixed once the curvature is derived from the entropy response.

To connect this expression with the conventional linear theory, write
\begin{align}
	\rho=\bar{\rho}(1+\delta).
\end{align}
Then
\begin{align}
	\ln\rho
	=
	\ln\bar{\rho}
	+
	\delta
	-
	\frac{1}{2}\delta^2
	+
	\frac{1}{3}\delta^3
	+\cdots ,
\end{align}
and hence
\begin{align}
	H_{ij}
	&=
	-\partial_i\partial_j\delta
	+
	\partial_i\delta\,\partial_j\delta
	+
	\delta\,\partial_i\partial_j\delta
	+O(\delta^3).
\end{align}
At leading order,
\begin{align}
	H_{ij}
	\simeq
	-\partial_i\partial_j\delta .
\end{align}
Thus the Hessian variable used in the conventional Gaussian peak theory is recovered as the linear approximation of the entropy-response curvature tensor.
In this sense, BBKS uses the Gaussian-linear limit of a more general logarithmic curvature.
The present formulation therefore does not simply reinterpret the standard Hessian.
It identifies the nonlinear geometric tensor whose leading-order limit reproduces the BBKS curvature variable and whose higher-order corrections are determined by the variational structure of entropy in Wasserstein space.

\subsection{Curvature Invariants and Positive Curvature Structure}

The curvature tensor $H_{ij}$ is a symmetric matrix, and the local geometry is characterized by its eigenvalue structure.
In particular, the trace and determinant are defined as
\begin{align}
	J_1=\mathrm{Tr}(H)
\end{align}
\begin{align}
	J_3=\det(H).
\end{align}
$J_1$ is interpreted as a measure of mean curvature, and $J_3$ as a measure of local volume contraction.
Let $\mathrm{Sym}(3)$ denote the vector space of real symmetric $3\times 3$ matrices.
The curvature tensor is an element $H\in\mathrm{Sym}(3)$, and its local shape is characterized by eigenvalues $\lambda_i$.

The positive curvature condition is equivalent to $H$ being positive definite, and is defined as the positive curvature cone
\begin{align}
	\mathcal{C}_+=\{H\in\mathrm{Sym}(3)\mid \bm{v}^\top H \bm{v}>0\ \text{for all}\ \bm{v}\neq 0\}.
\end{align}
This definition is equivalent to the eigenvalue condition
\begin{align}
	\lambda_i>0.
\end{align}
Furthermore, $\mathcal{C}_+$ is a convex set closed under positive scalar multiplication and forms a convex cone with the origin as its apex.
Thus, the peak condition is expressed as a geometric constraint that the curvature tensor belongs to this positive curvature cone.
In this sense, local maxima of the density distribution are understood not as merely analytic conditions, but as membership conditions in a specific region of curvature tensor space.

\subsection{General Definition of Stationary Points and Peaks}

In order to characterize the local structure of a distribution, it is necessary to consider both the gradient and curvature simultaneously.
In the previous subsections, the logarithmic curvature tensor was derived as the local response tensor of the entropy force induced by Wasserstein displacement perturbations.
The next step is to identify which geometric conditions correspond to local extrema of the density distribution.

Defining the entropy-gradient field as
\begin{align}
	g_i(\bm{x})=\partial_i\ln\rho(\bm{x}),
\end{align}
stationary points are given by
\begin{align}
	g=0.
\end{align}
Since $g_i$ is the gradient of the logarithmic density, this condition means that the local entropy force vanishes.
Therefore, stationary points correspond to equilibrium points of the entropy-gradient flow in the local measure geometry.

Furthermore, for such a point to be a local maximum, the curvature tensor must be positive definite,
\begin{align}
	H\in\mathcal{C}_+.
\end{align}
Therefore, a peak is defined as a point where the entropy-gradient field vanishes and the logarithmic curvature is positive in measure space.
This definition does not depend on a specific probability distribution and holds for general density fields.

The positivity condition may also be interpreted geometrically.
Since
\begin{align}
	H_{ij}=-\partial_i\partial_j\ln\rho,
\end{align}
the condition $H\in\mathcal{C}_+$ implies that the entropy force locally converges toward the stationary point.
Thus, a peak is not merely a local maximum of the density field itself, but a locally stable concentration point of the entropy-gradient flow.

Note that, given that the curvature tensor is defined as the Hessian of $\ln\rho$, the choice of $\partial_i\ln\rho$ as the gradient is natural.
Moreover, since the location of stationary points coincides for $\partial_i\rho=0$ and $\partial_i\ln\rho=0$, we adopt $g_i=\partial_i\ln\rho$ as the fundamental variable in the general theory.
In this sense, the basic variables $(\rho,g,H)$ in this work overlap with the set of local variables used in tensor bias theory, but their role here is not as arguments of a bias function, but as variables of a conditional measure.

This distinction is important conceptually.
In conventional local bias theories, tensor quantities are introduced as arguments of a response expansion.
In contrast, in the present formulation the local tensor variables themselves constitute the coordinates of a conditional geometric measure.
Thus, the central object is not a bias functional, but the measure structure defined on the local tensor space.

\subsection{Point Process Representation of Stationary Points}

In order to connect the above definitions from local geometric conditions to statistical quantities, it is necessary to represent the set of peaks as a point process in space.
In the classical theory of extrema statistics due to Rice, the number density of stationary points or extrema is described using the Jacobian associated with the zeros of the gradient field \citep{rice1944mathematical, rice1945mathematical}.
This idea was later introduced into cosmology for the statistics of maxima in Gaussian random fields, and was formulated in BBKS as the fundamental expression for peak statistics in three-dimensional fields \citep{1986ApJ...304...15B}.
In the present work, however, we reinterpret this structure geometrically as a measure induced by the logarithmic curvature tensor field.

Let $\bm{x}_p$ denote the peak positions.
Then the corresponding point process is expressed as
\begin{align}
	n_{\rm pk}(\bm{x})=\sum_p \deltadir^{(3)}(\bm{x}-\bm{x}_p)
\end{align}
where the sum runs over all points satisfying the stationary condition $g(\bm{x}_p)=0$ and the positive curvature condition $H(\bm{x}_p)\in\mathcal{C}_+$.

Expanding the gradient field $g_i(\bm{x})=\partial_i\ln\rho(\bm{x})$ around a stationary point $\bm{x}_p$, we obtain
\begin{align}
	g_i(\bm{x})
	=
	g_i(\bm{x}_p)
	+
	\partial_j g_i(\bm{x}_p)(\bm{x}-\bm{x}_p)_j
	+\cdots
\end{align}
At the stationary point, $g_i(\bm{x}_p)=0$, and furthermore
\begin{align}
	\partial_j g_i(\bm{x}_p)
	=
	\partial_j\partial_i\ln\rho(\bm{x}_p)
\end{align}
holds.
On the other hand, the curvature tensor introduced in this work is given by
\begin{align}
	H_{ij}(\bm{x})
	=
	-\partial_i\partial_j\ln\rho(\bm{x}).
\end{align}
Therefore, in the vicinity of a stationary point,
\begin{align}
	g_i(\bm{x})
	\simeq
	-H_{ij}(\bm{x}_p)(\bm{x}-\bm{x}_p)_j .
\end{align}
This relation is important because it shows that the local mapping between coordinate displacements and entropy-gradient displacements is governed by the logarithmic curvature tensor itself.

In other words, the curvature tensor acts as the local Jacobian of the entropy-gradient flow.
The determinant factor appearing in extrema statistics is therefore not merely a technical weight associated with coordinate transformation.
It represents the local volume deformation induced by the entropy-response geometry.

Using the change-of-variables formula for multivariate delta functions,
\begin{align}
	\deltadir^{(3)}(g(\bm{x}))
	=
	\sum_p
	\frac{
		\deltadir^{(3)}(\bm{x}-\bm{x}_p)
	}{
		\left|
			\det\left(\partial_j g_i\right)_{\bm{x}_p}
		\right|
	}
\end{align}
we obtain
\begin{align}
	\sum_p \deltadir^{(3)}(\bm{x}-\bm{x}_p)
	=
	\left|
		\det\left(\partial_j g_i\right)
	\right|
	\deltadir^{(3)}(g(\bm{x}))
\end{align}
\citep[e.g.,][]{adler1981geometry,1986ApJ...304...15B}.
To select only maxima, one imposes that the corresponding Hessian is negative definite.
In the present work, since we use $H_{ij}=-\partial_i\partial_j\ln\rho$, this condition is written as $H\in\mathcal{C}_+$.
Therefore, the peak point process is given by
\begin{align}
	n_{\rm pk}(\bm{x})
	=
	\left|
		\det\left(\partial_j g_i\right)
	\right|
	\deltadir^{(3)}(g(\bm{x}))
	\Theta(H(\bm{x})) .
\end{align}

Furthermore, since
\begin{align}
	\partial_j g_i=-H_{ij},
\end{align}
we can write the expression in the most compact form
\begin{align}
	n_{\rm pk}(\bm{x})
	=
	|\det H(\bm{x})|
	\deltadir^{(3)}(g(\bm{x}))
	\Theta(H(\bm{x})) .
\end{align}

This expression is one of the central results of the present formulation.
It shows that the logarithmic curvature tensor induced by entropy geometry directly generates the geometric measure of the peak point process.

Moreover, the locations of stationary points themselves coincide for $\partial_i\rho=0$ and $\partial_i\ln\rho=0$.
Therefore, in the geometric formulation of this work, the essential Jacobian is $|\det H|$, which appears as the natural volume element when counting the zeros of the gradient field of the logarithmic density.

This shows that the determinant factor of the Hessian appearing in conventional extrema statistics functions here as the Jacobian of the entropy-response curvature tensor.
In other words, counting peaks amounts to evaluating the measure pushed forward from the zero set of the entropy-gradient field through the local volume transformation induced by the logarithmic curvature tensor.

This reinterpretation is important conceptually.
In the conventional Kac--Rice or BBKS formulation, the determinant factor is usually introduced as a Jacobian associated with the coordinate transformation between field derivatives and spatial coordinates.
In the present formulation, the same factor acquires an additional geometric meaning:
it represents the local volume response induced by entropy-driven mass rearrangement.
Thus, the point process structure is not imposed externally, but emerges naturally from the local geometry of the entropy-gradient flow.

\subsection{Peak Statistics as a Curvature-Conditioned Point Process}

Based on the above point process representation, the distribution of peaks is formulated as a conditional measure on the curvature tensor field.
Taking the expectation value of the point process, we obtain
\begin{align}
	\bar{n}_{\rm pk}
	=
	\left\langle
		|\det H|\,
		\deltadir^{(3)}(g)\,
		\Theta(H)
	\right\rangle .
\end{align}
Here $g_i(\bm{x})=\partial_i\ln\rho(\bm{x})$ is the gradient of the logarithmic density, and the locations of stationary points coincide with those obtained from $\partial_i\rho=0$.

Writing this expectation in terms of the joint distribution of local variables, we have
\begin{align}
	\bar{n}_{\rm pk}
	=
	\int
	\pd\rho\,
	\pd^3 g\,
	\pd H\,
	|\det H|\,
	p(\rho,g,H)\,
	\deltadir^{(3)}(g)\,
	\Theta(H).
\end{align}
Thus, when the gradient of the logarithmic density is taken as a fundamental variable, the geometric Jacobian is naturally extracted as $|\det H|$.

The structure of this expression is important.
Peak statistics are decomposed into three independent ingredients:
\begin{align}
	\text{probability distribution}
	\times
	\text{curvature constraint}
	\times
	\text{geometric measure}.
\end{align}
The probability distribution is encoded in $p(\rho,g,H)$, the peak selection is imposed by $\Theta(H)$, and the induced geometric measure is represented by $|\det H|$.
This decomposition will remain valid even beyond the Gaussian limit discussed later.

In what follows, in order to make the correspondence with curvature geometry as transparent as possible, we adopt this formulation based on the logarithmic density gradient as the canonical form.
Then the peak number density is given by
\begin{align}
	n_{\rm pk}
	=
	\int
	\pd\rho\,
	\pd^3 g\,
	\pd H\,
	|\det H|\,
	p(\rho,g,H)\,
	\deltadir^{(3)}(g)\,
	\Theta(H).
\end{align}
Here $p(\rho,g,H)$ is the joint distribution of the density, the gradient of the logarithmic density, and the curvature tensor, $\Theta(H)$ is the indicator function imposing the constraint of the positive curvature cone, and $|\det H|$ represents the local volume element near a stationary point, appearing as a geometric weight.

This expression shows that peak statistics are not constructed as a quantity dependent on a specific random field, but rather as a general conditional measure defined on the curvature tensor field.
Thus, peaks are understood not as an analytic condition of density maxima, but as natural structures in curvature geometry.

Importantly, while this expression reduces to the BBKS peak statistics when a Gaussian distribution is assumed, its derivation itself does not rely on Gaussianity.
Therefore, the BBKS theory is positioned not as the starting point of the construction, but as a solvable Gaussian-linear limit of a more general curvature-conditioned point process.

This distinction is central to the present work.
The fundamental object is not a specific Gaussian integral formula, but the conditional geometric measure defined on local tensor variables themselves.
Gaussianity only specifies one particular closed form of the joint distribution.
The geometric skeleton consisting of the entropy-gradient condition, the positive curvature constraint, and the induced Jacobian measure remains meaningful beyond the Gaussian limit.

This viewpoint will also be important in the later discussion of nonlinear structure, non-Gaussianity, and the hierarchy of response functions.

\subsection{Connection to the Linear Regime and Density Fluctuations}

We finally verify how the general curvature formulation connects to the conventional description based on density fluctuations.
Expanding the density field around the mean density $\bar{\rho}$,
\begin{align}
	\rho=\bar{\rho}(1+\delta),
\end{align}
we obtain
\begin{align}
	\ln\rho
	=
	\ln\bar{\rho}
	+
	\delta
	-
	\frac{1}{2}\delta^2
	+
	O(\delta^3).
\end{align}
In the linear regime where the density fluctuation is sufficiently small,
\begin{align}
	\ln\rho\simeq\delta,
\end{align}
and therefore
\begin{align}
	H_{ij}\simeq-\partial_i\partial_j\delta .
\end{align}
Similarly,
\begin{align}
	g_i=\partial_i\ln\rho\simeq\partial_i\delta ,
\end{align}
so that in the linear Gaussian limit,
\begin{align}
	\eta_i=\partial_i\delta .
\end{align}

Thus, the stationary-point condition and the positivity condition on the logarithmic curvature tensor reduce to the conventional gradient-zero condition and the Hessian sign condition used in Gaussian peak theory.
In this limit, the curvature tensor introduced above coincides with the second-derivative tensor appearing in the BBKS formulation.

The important point, however, is that the interpretation is different.
In the conventional BBKS framework, the Hessian of $\delta$ is introduced directly as a local descriptor of extrema.
In the present formulation, the same tensor appears only as the leading-order approximation of the logarithmic entropy-response curvature derived from Wasserstein geometry.
Therefore, the standard Hessian formalism is recovered not as an independent starting point, but as the Gaussian-linear limit of a more general nonlinear curvature structure.

This observation clarifies the role of BBKS within the present theory.
The Gaussian peak theory is embedded into a broader transport-geometric framework in which the logarithmic curvature tensor provides the fundamental geometric object, while the BBKS variables emerge as its linearized limit.
Once this structure is adopted, the higher-order nonlinear corrections are no longer arbitrary additions, but are fixed systematically by the entropy-response geometry itself.

\section{Gaussian Closure of the Curvature-Conditioned Point Process}
\label{sec:peaks}

\subsection{Linear Gaussian Limit as a Solvable Closure}

In Section~\ref{sec:curvature_geometry}, peak statistics were formulated as a conditional geometric measure on the logarithmic curvature tensor field.
The essential structure of the theory was shown to consist of three elements:
the joint distribution of local variables, the positive-curvature constraint, and the geometric Jacobian induced by the entropy-response curvature tensor.
Importantly, none of these structures requires Gaussianity.

The role of the Gaussian assumption is therefore not to define the point process itself, but to provide a solvable closure of the joint distribution of local tensor variables.
In other words, the curvature-conditioned point process introduced in the previous section is completely general, whereas the Gaussian theory corresponds to the special case in which the associated conditional measure can be evaluated explicitly in closed form.

To obtain this solvable limit, we first introduce the linear regime.
Writing the density field around the mean density $\bar{\rho}$ as
\begin{align}
	\rho=\bar{\rho}(1+\delta),
\end{align}
we consider the regime where the density fluctuation $\delta$ is sufficiently small.
In this limit,
\begin{align}
	\ln\rho
	=
	\ln\bar{\rho}
	+
	\delta
	+
	O(\delta^2)
\end{align}
holds, and therefore the logarithmic curvature tensor reduces to
\begin{align}
	H_{ij}
	\simeq
	-\partial_i\partial_j\delta .
\end{align}

Similarly, in Section~\ref{sec:curvature_geometry}, the peak point process was formulated using the logarithmic density gradient
\begin{align}
	g_i=\partial_i\ln\rho .
\end{align}
In the linear regime,
\begin{align}
	g_i
	\simeq
	\partial_i\delta ,
\end{align}
so we introduce the notation
\begin{align}
	\eta_i\equiv\partial_i\delta .
\end{align}
Thus, in this section, $\eta_i$ is adopted as the linear Gaussian limit of $g_i$ used in the general nonlinear formulation.
The peak locations are then given by
\begin{align}
	\partial_i\delta=0 .
\end{align}

The important point here is that the geometric structure derived in the previous section remains unchanged.
The stationary-point condition, the positive-curvature selection, and the induced Jacobian structure are all preserved.
What changes is only that the logarithmic curvature tensor is approximated by the linear Hessian of the density fluctuation field.
From this viewpoint, the conventional BBKS theory is not the starting point of the construction.
Rather, it corresponds to the Gaussian-linear closure of a more general curvature-conditioned point process defined on local tensor space.
This perspective also clarifies the relation to recent local tensor bias theories.
The variables $(\delta,\eta_i,H_{ij})$ overlap with the local tensor variables appearing in perturbative bias expansions, but here they are treated not as arguments of a bias functional, but as jointly distributed random variables defining a conditional geometric measure.

Therefore, in the linear regime, the curvature-conditioned point process introduced in Section~\ref{sec:curvature_geometry} reduces to a formulation described by the gradient-zero condition and the Hessian condition of the density fluctuation field.
We now impose the assumption of a Gaussian random field and evaluate explicitly the induced conditional measure.

\subsection{Gaussian Random Field and Finite-Dimensional Closure of Local Tensor Variables}

We now impose the assumption that the density fluctuation field
$\delta(\bm{x})$ is Gaussian with zero mean.
Its statistics are then completely specified by the power spectrum
$P(k)$ \citep[e.g.,][]{1980lssu.book.....P, Peacock1999Cosmology, takeuchi2025physics},
through
\begin{align}
	\left\langle
		\tilde{\delta}(\bm{k})
		\tilde{\delta}(\bm{k}')
	\right\rangle
	=
	(2\pi)^3
	\deltadir^{(3)}
	(\bm{k}+\bm{k}')
	P(k).
\end{align}

Since derivatives are linear operators, the local variables
\begin{align}
	\delta,
	\qquad
	\eta_i=\partial_i\delta,
	\qquad
	H_{ij}=-\partial_i\partial_j\delta
\end{align}
also form a multivariate Gaussian system.
Consequently, the infinite-dimensional functional probability
distribution of the random field closes into a finite-dimensional
joint distribution of local tensor variables.

The corresponding covariance structure is completely determined by
the spectral moments
\begin{align}
	\sigma_j^2
	=
	\int_0^\infty
	\frac{k^2\pd k}{2\pi^2}
	P(k)k^{2j},
\end{align}
which characterize the variances of the density,
gradient, and curvature sectors.
In particular, rotational symmetry implies
\begin{align}
	\langle\eta_i\eta_j\rangle
	=
	\frac{\sigma_1^2}{3}\delta_{ij},
\end{align}
and
\begin{align}
	\langle H_{ij}H_{kl}\rangle
	=
	\frac{\sigma_2^2}{15}
	\left(
		\delta_{ij}\delta_{kl}
		+
		\delta_{ik}\delta_{jl}
		+
		\delta_{il}\delta_{jk}
	\right).
\end{align}

The essential consequence of the Gaussian assumption is therefore
not merely calculational simplicity.
Rather, it provides a solvable closure of the curvature-conditioned
point process, reducing the peak problem from a functional measure
on random fields to a finite-dimensional measure on local tensor
space.

\subsection{Dimensionless Curvature Variables and Recovery of the BBKS Formula}

Under the Gaussian closure introduced above, the curvature-conditioned
point process becomes analytically tractable.
The statistics of peaks are then conveniently described by a small set
of dimensionless variables.

We introduce the normalized density amplitude
\begin{align}
	\nu
	=
	\frac{\delta}{\sigma_0},
\end{align}
and the dimensionless curvature
\begin{align}
	x
	=
	-\frac{\nabla^2\delta}{\sigma_2}.
\end{align}
The variable $x$ corresponds to the trace of the curvature tensor and
represents the leading rotationally invariant measure of local curvature.
In fact,
\begin{align}
	x
	=
	\frac{\mathrm{Tr}(H)}{\sigma_2}
\end{align}
holds in the Gaussian-linear limit.

The joint statistics of $(\nu,x)$ are completely specified by the
correlation coefficient
\begin{align}
	\gamma
	=
	\frac{\sigma_1^2}{\sigma_0\sigma_2},
\end{align}
which measures the coupling between density amplitude and curvature.
By Gaussianity,
\begin{align}
	\langle\nu^2\rangle
	=
	\langle x^2\rangle
	=
	1,
	\qquad
	\langle\nu x\rangle
	=
	\gamma ,
\end{align}
and therefore
\begin{align}
	p(\nu,x)
	\propto
	\exp
	\left[
		-\frac{
			\nu^2
			-
			2\gamma\nu x
			+
			x^2
		}
		{2(1-\gamma^2)}
	\right].
	\label{eq:joint_distribution_nu_x}
\end{align}

The mean peak number density follows from the curvature-conditioned
measure derived in Section~\ref{sec:curvature_geometry},
\begin{align}
	n_{\rm pk}
	=
	\int
	\pd H\,
	|\det H|\,
	p(H,\nu\,|\,\eta=0)\,
	\Theta(H),
\end{align}
where $\Theta(H)$ imposes the positive-curvature condition.
The detailed reduction of this integral to eigenvalue space is presented in Appendix~\ref{app:eigenvalue_measure}.
The important point for the present discussion is that the Gaussian closure reduces the peak problem to a finite-dimensional integral completely determined by the spectral moments $(\sigma_0,\sigma_1,\sigma_2)$.

Introducing the characteristic curvature scale
\begin{align}
	R_\ast
	=
	\sqrt{3}
	\frac{\sigma_1}{\sigma_2},
\end{align}
the resulting peak number density becomes
\begin{align}
	n_{\rm pk}(\nu)
	=
	\frac{1}{(2\pi)^2R_\ast^3}
	e^{-\nu^2/2}
	G(\gamma,\gamma\nu),
	\label{eq:BBKS_recovered}
\end{align}
where
\begin{align}
	G(\gamma,w)
	=
	\int_0^\infty
	\pd x\,
	f(x)
	\frac{
		1
	}
	{
		\sqrt{2\pi(1-\gamma^2)}
	}
	\exp
	\left[
		-\frac{
			(x-w)^2
		}
		{
			2(1-\gamma^2)
		}
	\right].
\end{align}
The function $f(x)$ arises from the positive-curvature constraint and
its explicit form is identical to that obtained by BBKS
\citep{1986ApJ...304...15B}.

Equation~(\ref{eq:BBKS_recovered}) is therefore exactly the classical
BBKS peak abundance formula.
The interpretation, however, is different.
In the conventional formulation, Eq.~(\ref{eq:BBKS_recovered}) is taken
as the starting point of peak statistics.
In the present framework, it appears as the explicit evaluation of a
more general curvature-conditioned measure under Gaussian closure.

This result clarifies the position of BBKS within the transport-geometric
framework developed here.
The Gaussian theory does not define the geometric structure itself.
Rather, it corresponds to a solvable Gaussian-linear realization of the
underlying curvature-conditioned point process.

\subsection{First Nonlinear Correction from Logarithmic Curvature}\label{subsec:nonlinear_correction_1st}

One of the advantages of the present formulation is that nonlinear corrections are introduced in a geometrically controlled manner once the logarithmic curvature tensor is adopted.
From Section~\ref{sec:curvature_geometry}, the entropy-response curvature tensor is
\begin{align}
	H_{ij}
	=
	-\partial_i\partial_j\ln\rho.
\end{align}
Expanding around
\begin{align}
	\rho=\bar{\rho}(1+\delta),
\end{align}
we obtained
\begin{align}
	H_{ij}
	=
	-\partial_i\partial_j\delta
	+
	\partial_i\delta\,\partial_j\delta
	+
	\delta\,\partial_i\partial_j\delta
	+
	O(\delta^3).
\end{align}

We now decompose the curvature tensor into the Gaussian-linear part and the first nonlinear correction:
\begin{align}
	H_{ij}
	=
	H_{ij}^{(0)}
	+
	H_{ij}^{(2)}
	+
	O(\delta^3),
\end{align}
with
\begin{align}
	H_{ij}^{(0)}
	=
	-\partial_i\partial_j\delta,
\end{align}
and
\begin{align}
	H_{ij}^{(2)}
	=
	\partial_i\delta\,\partial_j\delta
	-
	\delta H_{ij}^{(0)}.
\end{align}
The first term in $H_{ij}^{(2)}$ is positive semi-definite and enhances curvature along strong gradient directions.
The second term rescales the local Hessian amplitude according to the local density fluctuation itself.
Thus, nonlinear curvature corrections are intrinsically anisotropic and depend on both local slope and local density contrast.

To quantify the deformation of local peak geometry, let
\begin{align}
	H_{ij}^{(0)} e_a^j
	=
	\lambda_a^{(0)} e_a^i
\end{align}
be the eigenvalue system of the Gaussian-linear curvature tensor.
To first order in the nonlinear correction, the shifted eigenvalues become
\begin{align}
	\lambda_a
	=
	\lambda_a^{(0)}
	+
	\Delta\lambda_a
	+
	O(\delta^3),
\end{align}
where
\begin{align}
	\Delta\lambda_a
	=
	e_a^i
	H_{ij}^{(2)}
	e_a^j.
\end{align}
Therefore,
\begin{align}
	\Delta\lambda_a
	=
	(e_a\cdot\nabla\delta)^2
	-
	\delta\lambda_a^{(0)}.
\end{align}
This expression gives the first nonlinear deformation of the principal curvatures induced by the entropy-response geometry.

Several qualitative consequences follow immediately.
\begin{itemize}

\item
Curvature enhancement becomes direction dependent through the projected gradient term $(e_a\cdot\nabla\delta)^2$.
Thus, nonlinear peaks are no longer characterized solely by the Hessian amplitude, but also by the anisotropy of local density gradients.

\item
High-density regions with positive $\delta$ tend to amplify the magnitude of already-positive principal curvatures through the second term.
Therefore, concentrated nonlinear structures are geometrically sharpened relative to the Gaussian prediction.

\item
Since the determinant factor in the point-process measure depends on the product of principal curvatures,
\begin{align}
	|\det H|
	=
	\prod_a |\lambda_a|,
\end{align}
the nonlinear correction modifies the peak abundance itself through shifts of the local curvature spectrum.

\end{itemize}
These effects are absent in the conventional Gaussian BBKS theory because they originate from the logarithmic entropy-response structure of the curvature tensor.
In the present formulation, however, they arise automatically once the geometric origin of curvature is specified.
This illustrates how the curvature-conditioned point process naturally extends beyond the Gaussian-linear approximation while preserving the underlying geometric structure.

\subsection{Determinant Correction to the Peak Measure}
\label{subsec:determinant_peak_measure}

The eigenvalue shift derived above shows how the principal curvatures are deformed by the logarithmic entropy-response structure. We now translate this deformation into the actual measure factor that enters the peak point process. This step is important because the peak abundance is not controlled only by the location of stationary points or by the positivity of the curvature tensor, but also by the Jacobian factor $|\det H|$. Therefore, a nonlinear deformation of the curvature tensor directly induces a nonlinear correction to the peak measure.

We start from the decomposition
\begin{align}
	H_{ij}
	=
	H_{ij}^{(0)}
	+
	H_{ij}^{(2)}
	+
	O(\delta^3),
	\label{eq:H_decomposition_for_det}
\end{align}
where
\begin{align}
	H_{ij}^{(0)}
	=
	-\partial_i\partial_j\delta,
	\qquad
	H_{ij}^{(2)}
	=
	\partial_i\delta\,\partial_j\delta
	-
	\delta H_{ij}^{(0)}.
	\label{eq:H0_H2_for_det}
\end{align}
Here $H^{(0)}=O(\delta)$ and $H^{(2)}=O(\delta^2)$ in the formal amplitude expansion. Assuming that $H^{(0)}$ is nondegenerate at the stationary point under consideration, we can expand the determinant as
\begin{align}
	\det H
	&=
	\det\left(H^{(0)}+H^{(2)}\right)
	+
	O(\delta^5)
	\nonumber\\
	&=
	\det H^{(0)}
	\det\left[
		I+\left(H^{(0)}\right)^{-1}H^{(2)}
	\right]
	+
	O(\delta^5).
	\label{eq:det_expansion_start}
\end{align}
Since $\det H^{(0)}=O(\delta^3)$ and $\left(H^{(0)}\right)^{-1}H^{(2)}=O(\delta)$, retaining the first correction to the determinant gives
\begin{align}
	\det H
	=
	\det H^{(0)}
	\left[
		1+
		\mathrm{Tr}\left(
			\left(H^{(0)}\right)^{-1}H^{(2)}
		\right)
	\right]
	+
	O(\delta^5).
	\label{eq:det_expansion_trace}
\end{align}
Equivalently, in the eigenbasis of $H^{(0)}$, where
\begin{align}
	H_{ij}^{(0)}e_a^j
	=
	\lambda_a^{(0)}e_a^i,
\end{align}
we have
\begin{align}
	\det H
	=
	\prod_a
	\left(
		\lambda_a^{(0)}+\Delta\lambda_a
	\right)
	+
	O(\delta^5),
\end{align}
and hence
\begin{align}
	\det H
	=
	\det H^{(0)}
	\left[
		1+
		\sum_a
		\frac{\Delta\lambda_a}{\lambda_a^{(0)}}
	\right]
	+
	O(\delta^5).
	\label{eq:det_expansion_eigen}
\end{align}
Using
\begin{align}
	\Delta\lambda_a
	=
	(e_a\cdot\nabla\delta)^2
	-
	\delta\lambda_a^{(0)},
\end{align}
we obtain
\begin{align}
	\sum_a
	\frac{\Delta\lambda_a}{\lambda_a^{(0)}}
	=
	\sum_a
	\frac{(e_a\cdot\nabla\delta)^2}{\lambda_a^{(0)}}
	-
	3\delta.
\end{align}
Therefore, the determinant factor becomes
\begin{align}
	\det H
	=
	\det H^{(0)}
	\left[
		1
		-
		3\delta
		+
		\sum_a
		\frac{(e_a\cdot\nabla\delta)^2}{\lambda_a^{(0)}}
	\right]
	+
	O(\delta^5).
	\label{eq:det_H_nonlinear_correction}
\end{align}
This is the first explicit nonlinear correction to the geometric Jacobian of the peak point process induced by logarithmic curvature.

The same result can be written in coordinate-free form as
\begin{align}
	\det H
	=
	\det H^{(0)}
	\left[
		1
		+
		\mathrm{Tr}
		\left(
			\left(H^{(0)}\right)^{-1}
			\nabla\delta\,\nabla\delta^{\top}
		\right)
		-
		3\delta
	\right]
	+
	O(\delta^5),
	\label{eq:det_H_coordinate_free}
\end{align}
where $(\nabla\delta\,\nabla\delta^{\top})_{ij}=\partial_i\delta\,\partial_j\delta$. This expression separates the correction into an isotropic amplitude term, $-3\delta$, and an anisotropic gradient term controlled by the inverse of the Gaussian-linear curvature tensor. The latter term measures how strongly the local density gradient is aligned with soft or hard curvature directions. In particular, directions with small $\lambda_a^{(0)}$ are more sensitive to the projected gradient contribution.

\begin{figure}[t]
	\centering
	\includegraphics[width=0.7\textwidth]{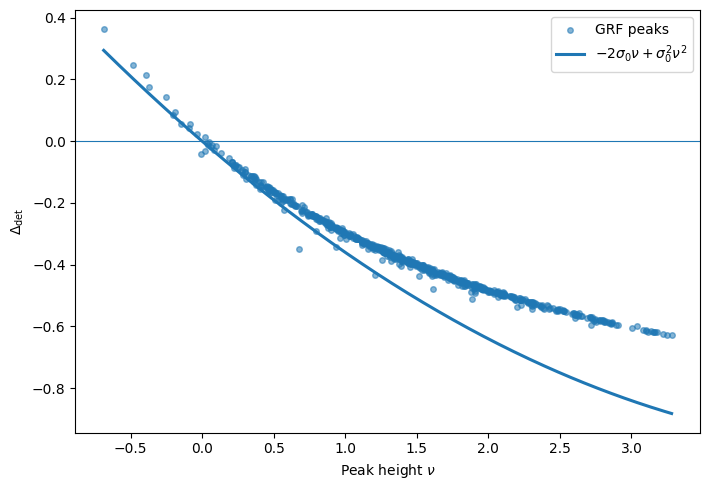}
	\caption{
	Realization-level test of the logarithmic entropy-response curvature in a two-dimensional Gaussian random field.
	The determinant deformation
	$\Delta_{\det}$ defined in Eq.~\eqref{eq:determinant_relative_deformation}
	evaluated on the corresponding peak ensemble as a function of peak height $\nu$.
	The solid curve shows the leading-order theoretical prediction
	$-2\sigma_0\nu+\sigma_0^2\nu^2$
	for the determinant-sector deformation in two dimensions.
	The numerical realization exhibits the predicted systematic deformation of the curvature-conditioned point-process measure induced by the logarithmic curvature tensor.
    The residual deviation at high peak height is expected from the perturbative nature of the present approximation and anticipates the cone-boundary effects discussed later in Sec.~\ref{subsec:cone_boundary_effects}.
    }
	\label{fig:grf_determinant_deformation}
\end{figure}

We now insert this expansion into the curvature-conditioned peak measure. To leading Gaussian-linear order, the peak measure is
\begin{align}
	n_{\rm pk}^{(0)}(\bm{x})
	=
	\left|
		\det H^{(0)}(\bm{x})
	\right|
	\deltadir^{(3)}(\eta(\bm{x}))
	\Theta\!\left(H^{(0)}(\bm{x})\right),
	\label{eq:npk_linear_measure}
\end{align}
where $\eta_i=\partial_i\delta$. The logarithmic curvature correction changes the Jacobian factor, the peak-selection domain, and, in principle, the gradient-zero condition if one works beyond the linear approximation to $g_i=\partial_i\ln\rho$. In the present subsection, we isolate the correction to the geometric Jacobian while keeping the Gaussian stationary-point condition fixed. This gives the first controlled deformation of the BBKS peak measure:
\begin{align}
	n_{\rm pk}(\bm{x})
	=
	n_{\rm pk}^{(0)}(\bm{x})
	+
	\Delta n_{\rm pk}^{\det}(\bm{x})
	+
	\cdots ,
	\label{eq:npk_expansion}
\end{align}
with
\begin{align}
	\Delta n_{\rm pk}^{\det}(\bm{x})
	=
	\left|
		\det H^{(0)}
	\right|
	\left[
		-3\delta
		+
		\mathrm{Tr}
		\left(
			\left(H^{(0)}\right)^{-1}
			\nabla\delta\,\nabla\delta^{\top}
		\right)
	\right]
	\deltadir^{(3)}(\eta)
	\Theta\!\left(H^{(0)}\right).
	\label{eq:delta_npk_det}
\end{align}
Equivalently, in the eigenbasis of $H^{(0)}$,
\begin{align}
	\Delta n_{\rm pk}^{\det}(\bm{x})
	=
	\left|
		\det H^{(0)}
	\right|
	\left[
		-3\delta
		+
		\sum_a
		\frac{(e_a\cdot\nabla\delta)^2}{\lambda_a^{(0)}}
	\right]
	\deltadir^{(3)}(\eta)
	\Theta\!\left(H^{(0)}\right).
	\label{eq:delta_npk_det_eigen}
\end{align}
Equations~(\ref{eq:delta_npk_det}) and (\ref{eq:delta_npk_det_eigen}) are useful because they display explicitly how the entropy-response curvature modifies the BBKS point-process weight.
To verify explicitly that the logarithmic entropy-response curvature produces a measurable deformation of the peak measure, we generated two-dimensional Gaussian random field realizations and compared the standard Gaussian-linear curvature tensor
\begin{align}
	H_{ij}^{(0)}
	=
	-\sigma_0\,\partial_i\partial_j\delta_R
\end{align}
with the nonlinear logarithmic curvature tensor
\begin{align}
	H_{ij}^{\log}
	=
	-\partial_i\partial_j
	\ln(1+\sigma_0\delta_R),
	\label{eq:log_curvature_grf_test}
\end{align}
where $\delta_R$ denotes a smoothed Gaussian random field.
The corresponding determinant deformation is quantified by
\begin{align}
	\Delta_{\det}
	=
	\frac{
		|\det H^{\log}|
		-
		|\det H^{(0)}|
	}{
		|\det H^{(0)}|
	}.
	\label{eq:determinant_relative_deformation}
\end{align}
Figure~\ref{fig:grf_determinant_deformation} shows the resulting realization-level deformation of the curvature-conditioned peak measure.
The residual deviation at high peak height is likely associated with higher-order nonlinear corrections and boundary effects of the positive-curvature cone, which become increasingly important for strongly concentrated peaks.

There is an important simplification at exact stationary points of the linear field. 
Since the delta function $\deltadir^{(3)}(\eta)$ enforces $\eta_i=\partial_i\delta=0$, the projected gradient term vanishes inside expectation values involving only smooth test functions of the local variables. 
This simplification is performed within a fixed linear stationary-point expansion.
More precisely, we evaluate the determinant deformation on the stationary set of the Gaussian-linear field,
\begin{align}
	\eta_i=\partial_i\delta=0,
\end{align}
while the difference between the nonlinear stationary condition
\begin{align}
	g_i=\partial_i\ln\rho=0
\end{align}
and the linear condition is treated as part of the higher-order correction
$\Delta n_{\rm pk}^{g}$ discussed below.
Therefore, Eq.~(3.69) is valid consistently up to the order at which the stationary-point displacement itself is neglected.
In that restricted case, the determinant correction reduces to
\begin{align}
	\Delta n_{\rm pk}^{\det}(\bm{x})
	=
	-3\delta\,
	\left|
		\det H^{(0)}
	\right|
	\deltadir^{(3)}(\eta)
	\Theta\!\left(H^{(0)}\right),
	\label{eq:delta_npk_det_stationary}
\end{align}
up to corrections associated with the deformation of the peak-selection domain and the nonlinear stationary condition. Thus, the leading entropy-curvature correction to the determinant measure at fixed Gaussian stationary points is an amplitude-dependent rescaling of the BBKS Jacobian. This result is already nontrivial: even before including the displacement of peak positions or changes of the positive-curvature boundary, the logarithmic curvature changes the peak weight by a factor proportional to the local density contrast.

Taking the ensemble average gives the corresponding correction to the peak number density:
\begin{align}
	\Delta n_{\rm pk}^{\det}
	=
	-3
	\left\langle
		\delta\,
		\left|
			\det H^{(0)}
		\right|
		\deltadir^{(3)}(\eta)
		\Theta\!\left(H^{(0)}\right)
	\right\rangle .
	\label{eq:delta_npk_det_average}
\end{align}
For peaks of fixed height $\nu=\delta/\sigma_0$, this becomes
\begin{align}
	\Delta n_{\rm pk}^{\det}(\nu)
	=
	-3\sigma_0\nu\,
	n_{\rm pk}^{(0)}(\nu),
	\label{eq:delta_npk_det_fixed_nu_simple}
\end{align}
again within the restricted approximation in which only the determinant factor is corrected. More generally, without fixing $\nu$, the correction can be written as
\begin{align}
	\Delta n_{\rm pk}^{\det}
	=
	-3
	\int
	\pd\nu\,
	\sigma_0\nu\,
	n_{\rm pk}^{(0)}(\nu).
	\label{eq:delta_npk_det_integrated}
\end{align}
These expressions show explicitly how the logarithmic curvature produces a calculable deformation of the BBKS peak abundance.

It should be emphasized that Eqs.~(\ref{eq:delta_npk_det})--(\ref{eq:delta_npk_det_fixed_nu_simple}) represent only the Jacobian part of the full nonlinear correction. The complete correction also contains the deformation of the stationary condition,
\begin{align}
	g_i
	=
	\partial_i\ln\rho
	=
	\partial_i\delta
	-
	\delta\,\partial_i\delta
	+
	O(\delta^3),
	\label{eq:gradient_nonlinear_expansion}
\end{align}
and the deformation of the curvature-selection condition,
\begin{align}
	\Theta(H)
	=
	\Theta\!\left(H^{(0)}+H^{(2)}\right).
\end{align}
Formally, the latter can be written as
\begin{align}
	\Theta\!\left(H^{(0)}+H^{(2)}\right)
	=
	\Theta\!\left(H^{(0)}\right)
	+
	\delta_{\partial\mathcal{C}_+}\!\left(H^{(0)}\right)
	\left[
		H^{(2)}
	\right]
	+
	\cdots ,
	\label{eq:theta_cone_deformation}
\end{align}
where $\delta_{\partial\mathcal{C}_+}$ denotes the distribution supported on the boundary of the positive-curvature cone. This boundary contribution represents the shift of configurations close to zero principal curvature across the peak-selection threshold. Although this term is usually subleading for well-conditioned peaks with all $\lambda_a^{(0)}>0$, it is conceptually important because it shows that the nonlinear curvature affects not only the weight of already-selected peaks, but also the selection domain itself.

Combining these ingredients, the full nonlinear deformation of the peak point process can be organized schematically as
\begin{align}
	n_{\rm pk}
	=
	n_{\rm pk}^{(0)}
	+
	\Delta n_{\rm pk}^{\det}
	+
	\Delta n_{\rm pk}^{g}
	+
	\Delta n_{\rm pk}^{\Theta}
	+
	\cdots ,
	\label{eq:full_npk_deformation_schematic}
\end{align}
where $\Delta n_{\rm pk}^{\det}$ is the determinant correction derived explicitly above, $\Delta n_{\rm pk}^{g}$ is generated by the nonlinear deformation of the stationary condition, and $\Delta n_{\rm pk}^{\Theta}$ comes from the shift of the positive-curvature cone. Thus, the logarithmic entropy-response curvature does not merely reinterpret the Gaussian BBKS formula. It provides a systematic expansion of the peak point-process measure beyond the Gaussian-linear limit, with each correction term having a definite geometric origin.
We should note that there is a significant deviation between the numerical result and the theoretical curve. 
We will revisit this issue in Section~\ref{subsec:cone_boundary_effects}

\section{Nonlinear Deformation of Curvature Statistics}
\label{sec:nonlinear_deformation}

Section~\ref{sec:peaks} showed that the logarithmic entropy-response curvature induces explicit nonlinear corrections to the curvature-conditioned point process.
In particular, the determinant expansion derived there already demonstrated that the BBKS peak measure acquires calculable nonlinear deformations beyond the Gaussian-linear approximation.
In this section, we develop these nonlinear effects systematically at the level of the curvature spectrum, the geometric Jacobian, and the associated peak abundance.

\subsection{Nonlinear Deformation of the Curvature Spectrum}
\label{subsec:nonlinear_curvature_spectrum}

Using the logarithmic curvature expansion derived in Section~\ref{subsec:nonlinear_correction_1st}, the leading nonlinear shift of the principal curvature spectrum is
\begin{align}
	\Delta\lambda_a
	=
	(e_a\cdot\nabla\delta)^2
	-
	\delta\lambda_a^{(0)}.
	\label{eq:nonlinear_eigenvalue_shift}
\end{align}

Equation~\eqref{eq:nonlinear_eigenvalue_shift} is the first explicit nonlinear consequence of the entropy-response curvature.
It shows that the curvature spectrum is not merely rescaled by the density amplitude.
It is deformed anisotropically by the projected gradient of the density field along each principal curvature direction.
Thus, even if two points have the same linear Hessian spectrum, their nonlinear curvature spectra differ whenever their density gradients have different orientations relative to the eigenvectors of $H^{(0)}$.

This is a genuine extension beyond the Gaussian-linear peak theory.
In the Gaussian closure, the local peak shape is determined by the eigenvalues of the Hessian alone.
In the logarithmic curvature theory, by contrast, the nonlinear correction couples the curvature eigenframe to the gradient direction.
The local peak geometry is therefore controlled not only by the amplitude and Hessian of the density fluctuation, but also by the relative orientation between the density-gradient vector and the principal curvature axes.

The first term in Eq.~\eqref{eq:nonlinear_eigenvalue_shift} is positive semi-definite.
It enhances the principal curvature in directions along which the density gradient has a large projection.
The second term is an amplitude-dependent rescaling of the Gaussian-linear curvature.
For positive $\delta$, this term reduces the eigenvalue shift if $\lambda_a^{(0)}>0$, whereas for negative $\delta$ it enhances it.
Therefore, the nonlinear curvature spectrum reflects a competition between directional gradient focusing and local amplitude modulation.

This result also clarifies why the logarithmic curvature tensor is not simply a change of variables applied to the BBKS Hessian.
The nonlinear terms produce new tensorial information that is absent from the linear Hessian spectrum.
In particular, the projected-gradient contribution depends on the eigenvectors of the curvature tensor and not only on rotational invariants such as the trace or determinant.
Hence the nonlinear curvature theory naturally introduces orientation-dependent peak deformation.

\subsection{Determinant Deformation and the Peak Measure}
\label{subsec:determinant_deformation}

The peak point process is weighted by the determinant of the curvature tensor.
Consequently, the nonlinear deformation of $H$ induces a direct deformation of the peak measure.
This step is important because the observable peak abundance is not determined only by the positivity of the principal curvatures, but also by the local volume factor associated with the curvature map.

For a nondegenerate stationary point, write
\begin{align}
	H=H^{(0)}+H^{(2)}+O(\delta^3).
\end{align}
Assuming that $H^{(0)}$ is invertible, the determinant is expanded as
\begin{align}
	\det H
	=
	\det H^{(0)}
	\det\left[
		I+(H^{(0)})^{-1}H^{(2)}
	\right].
\end{align}
To first order in the nonlinear correction,
\begin{align}
	\det H
	=
	\det H^{(0)}
	\left\{
		1+
		\Tr\left[(H^{(0)})^{-1}H^{(2)}\right]
	\right\}
	+
	O\!\left((H^{(2)})^2\right).
	\label{eq:determinant_expansion_general}
\end{align}
In the eigenbasis of $H^{(0)}$, this becomes
\begin{align}
	\det H
	=
	\det H^{(0)}
	\left[
		1+
		\sum_{a=1}^3
		\frac{\Delta\lambda_a}{\lambda_a^{(0)}}
	\right]
	+
	O\!\left((H^{(2)})^2\right).
	\label{eq:determinant_expansion_eigen}
\end{align}
Substituting Eq.~\eqref{eq:nonlinear_eigenvalue_shift}, we obtain
\begin{align}
	\det H
	=
	\det H^{(0)}
	\left[
		1+
		\sum_{a=1}^3
		\frac{(e_a\cdot\nabla\delta)^2}{\lambda_a^{(0)}}
		-
		3\delta
	\right]
	+
	O\!\left((H^{(2)})^2\right).
	\label{eq:determinant_deformation}
\end{align}

Equation~\eqref{eq:determinant_deformation} gives the nonlinear deformation of the geometric measure entering peak statistics.
The first correction term is directional and depends on the projection of the density gradient onto each principal curvature axis.
The second correction term is isotropic and depends only on the local density amplitude.
Thus, the determinant weight separates into an anisotropic gradient-focusing contribution and an isotropic amplitude-modulation contribution.

This result is operationally important.
The determinant factor is the Jacobian that converts the zero set of the gradient field into a spatial point process.
Therefore, Eq.~\eqref{eq:determinant_deformation} shows that the nonlinear entropy-response curvature modifies the point-process measure itself.
The effect is not a postulated correction to the BBKS formula, but follows directly from the logarithmic curvature structure.

It is also worth noting that the correction becomes large when one of the principal curvatures is small.
This is natural from the viewpoint of curvature geometry.
Near a nearly degenerate stationary point, a small nonlinear perturbation can significantly change the local volume factor.
Therefore, nonlinear curvature effects are expected to be especially important near the boundary of the positive curvature cone.

\subsection{Anisotropic Enhancement of Local Curvature}
\label{subsec:anisotropic_enhancement}

The determinant deformation derived above shows that nonlinear curvature corrections are intrinsically anisotropic.
This anisotropy is controlled by the quantities $	(e_a\cdot\nabla\delta)^2$ which measure the squared components of the density gradient along the principal curvature directions.
These terms are absent in the Gaussian-linear theory because there the local shape is fully determined by the Hessian spectrum.
In the logarithmic curvature theory, however, the eigenframe of the curvature tensor interacts with the local direction of density variation.

To make this point explicit, define the directional nonlinear enhancement factor
\begin{align}
	A_a
	\equiv
	\frac{(e_a\cdot\nabla\delta)^2}{\lambda_a^{(0)}}.
	\label{eq:directional_enhancement_factor}
\end{align}
Then the determinant deformation can be written as
\begin{align}
	\frac{\det H-\det H^{(0)}}{\det H^{(0)}}
	=
	\sum_{a=1}^3 A_a
	-
	3\delta
	+
	O\!\left((H^{(2)})^2\right).
	\label{eq:relative_det_deformation}
\end{align}
The quantity $A_a$ is large when the gradient has a strong projection along a direction with small linear curvature.
Thus, nonlinear corrections preferentially enhance directions that are weakly curved in the Gaussian-linear approximation but have substantial local density variation.

This behavior has a simple geometric interpretation.
The logarithmic curvature does not measure the bending of the density field alone.
It measures the response of the entropy force associated with the local mass distribution.
A strong gradient corresponds to rapid local rearrangement of mass, and the term $(e_a\cdot\nabla\delta)^2$ represents the focusing of this rearrangement along the $a$-th principal direction.
Thus, nonlinear peak geometry is sensitive not only to how sharply the field bends, but also to how mass is locally transported toward or away from the stationary region.

This anisotropic enhancement distinguishes the present formulation from a purely scalar nonlinear correction.
A scalar correction would modify the peak abundance only through invariants such as $\delta$ or $\Tr H$.
By contrast, Eq.~\eqref{eq:directional_enhancement_factor} depends on the relative orientation between $\nabla\delta$ and the eigenvectors of $H^{(0)}$.
Therefore, the logarithmic curvature framework predicts a directional deformation of local peak shapes.
This provides a concrete tensorial effect that is not visible in the standard Gaussian BBKS formula.

\subsection{Nonlinear Shift of Peak Abundance}
\label{subsec:nonlinear_peak_abundance}

We now translate the determinant deformation into a correction to the mean peak abundance.
Let $n_{\rm pk}^{(0)}$ denote the Gaussian-linear peak number density obtained from the curvature-conditioned point process, and write the nonlinear result as
\begin{align}
	n_{\rm pk}
	=
	n_{\rm pk}^{(0)}
	+
	\Delta n_{\rm pk}^{(2)}
	+
	\cdots .
\end{align}
The leading correction arises from two sources.
First, the determinant measure is deformed as shown in Eq.~\eqref{eq:determinant_deformation}.
Second, the joint distribution of local variables is itself modified once nonlinear curvature variables are used.
In this subsection, we isolate the first effect, namely the direct deformation of the geometric peak measure.

\begin{figure}[t]
	\centering
	\includegraphics[width=0.7\textwidth]{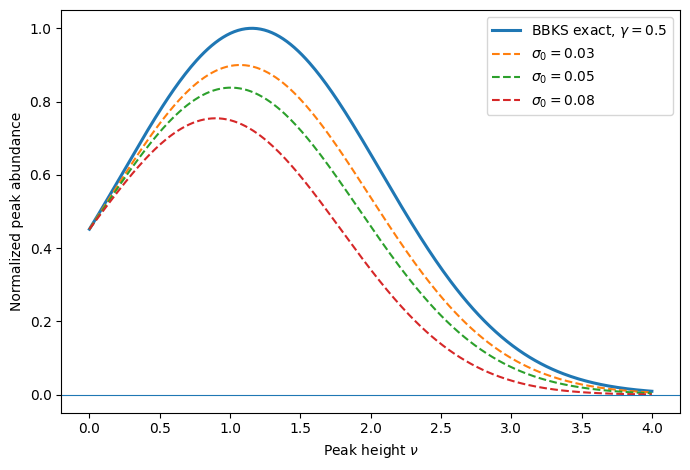}
	\caption{
	Relative nonlinear deformation of the BBKS peak abundance induced by the determinant correction to the logarithmic entropy-response curvature.
	The plotted curves correspond to the correction term derived from Eq.~\eqref{eq:determinant_peak_abundance_shift}, evaluated relative to the Gaussian BBKS reference abundance.
	The solid curve shows the exact BBKS prediction, while the dashed curves show the determinant-sector logarithmic-curvature correction for representative values of $\sigma_0$.
	The nonlinear correction suppresses the abundance of high peaks and becomes increasingly important as the fluctuation amplitude grows.
	}
	\label{fig:nonlinear_peak_shift}
\end{figure}

The measure-induced contribution is obtained by inserting the determinant expansion into the peak point process and evaluating it with the Gaussian-linear conditional measure.
This gives
\begin{align}
	\Delta n_{\rm pk,det}^{(2)}
	=
	\left\langle
		|\det H^{(0)}|
		\left[
			\sum_{a=1}^3
			\frac{(e_a\cdot\nabla\delta)^2}{\lambda_a^{(0)}}
			-
			3\delta
		\right]
		\delta_{\rm D}^{(3)}(\eta)
		\Theta(H^{(0)})
	\right\rangle_G,
	\label{eq:determinant_peak_abundance_shift}
\end{align}
where $\langle\cdots\rangle_G$ denotes the expectation value with respect to the Gaussian-linear joint distribution.
This effect is shown in Fig.~\ref{fig:nonlinear_peak_shift}. 
Equivalently, in terms of the peak-conditioned average,
\begin{align}
	\frac{\Delta n_{\rm pk,det}^{(2)}}{n_{\rm pk}^{(0)}}
	=
	\left\langle
		\sum_{a=1}^3
		\frac{(e_a\cdot\nabla\delta)^2}{\lambda_a^{(0)}}
		-
		3\delta
	\right\rangle_{{\rm pk},G}.
	\label{eq:relative_peak_abundance_shift}
\end{align}
Here $\langle\cdots\rangle_{{\rm pk},G}$ denotes averaging over the Gaussian peak ensemble, including the determinant weight and the positive curvature condition.
Geometrically, the suppression of high peaks reflects the fact that the logarithmic curvature sharpens concentrated structures through the entropy-response geometry.
As the local curvature increases, the associated Jacobian volume element contracts, reducing the effective configuration-space volume contributing to the peak measure.

Equation~\eqref{eq:relative_peak_abundance_shift} is a compact expression for the nonlinear shift of peak abundance induced by the logarithmic curvature measure.
It shows that the abundance correction is controlled by two competing effects.
The first is the anisotropic gradient-focusing contribution, which is positive when projected gradients are nonzero.
The second is the amplitude-modulation contribution, proportional to $-3\delta$.
The relative importance of these terms depends on the peak height, the curvature spectrum, and the alignment between the gradient field and the curvature eigenframe.

At exact Gaussian-linear peaks, the linear gradient variable satisfies $\eta=0$.
However, Eq.~\eqref{eq:relative_peak_abundance_shift} should not be read as a trivial vanishing statement.
The nonlinear correction is evaluated as a perturbation of the curvature-conditioned measure, and the gradient-square term represents the deformation of the logarithmic curvature away from the strictly linear peak condition.
Equivalently, it quantifies how the peak measure changes when the local entropy-response curvature is used instead of the linear Hessian.
This is precisely the regime in which nonlinear structure formation begins to deform the Gaussian peak ensemble.

The expression above provides a direct route to quantitative evaluation.
Given a smoothing scale and a Gaussian reference spectrum, the peak-conditioned average in Eq.~\eqref{eq:relative_peak_abundance_shift} can be evaluated either analytically under additional approximations or numerically by sampling the joint distribution of local tensor variables.
Thus, the nonlinear curvature theory yields a calculable correction to peak abundance rather than only a formal reinterpretation of the BBKS structure.

\subsection{Coarse-Grained Density and Scale-Dependent Curvature}
\label{subsec:coarse_grained_curvature}

Let $W_R$ be a smoothing kernel with characteristic scale $R$.
For a density field $\rho(\bm{x})$, the coarse-grained density is defined by
\begin{align}
	\rho_R(\bm{x})
	=
	\int \pd^3 y\,
	W_R(|\bm{x}-\bm{y}|)\rho(\bm{y}).
\end{align}
The scale-dependent logarithmic curvature tensor is then defined by
\begin{align}
	H_{ij}(\bm{x};R)
	=
	-\partial_i\partial_j\ln\rho_R(\bm{x}).
	\label{eq:scale_dependent_log_curvature}
\end{align}
The corresponding scale-dependent entropy-gradient field is
\begin{align}
	g_i(\bm{x};R)
	=
	\partial_i\ln\rho_R(\bm{x}).
\end{align}
Thus, at each scale $R$, peaks are selected by
\begin{align}
	g_i(\bm{x};R)=0,
	\qquad
	H(\bm{x};R)\in\mathcal{C}_+.
\end{align}

This definition makes clear that scale dependence enters before any statistical averaging is performed.
The local tensor variables themselves depend on $R$.
Therefore, changing the smoothing scale does not merely change the numerical value of the peak abundance.
It changes the map from the density field to the local curvature tensor field.

In the linear regime, where $\rho_R=\bar{\rho}(1+\delta_R)$ and $\ln\rho_R\simeq\ln\bar{\rho}+\delta_R$, Eq.~\eqref{eq:scale_dependent_log_curvature} reduces to
\begin{align}
	H_{ij}(\bm{x};R)
	\simeq
	-\partial_i\partial_j\delta_R(\bm{x}).
\end{align}
In this limit, the effect of smoothing is summarized by the scale-dependent spectral moments
\begin{align}
	\sigma_j^2(R)
	=
	\int_0^\infty
	\frac{k^2\pd k}{2\pi^2}
	P(k)k^{2j}W^2(kR).
	\label{eq:scale_dependent_spectral_moments}
\end{align}
However, beyond the linear regime, smoothing and taking the logarithm do not commute.
Consequently, the scale dependence of $H_{ij}(\bm{x};R)$ contains information that cannot be reduced to the scale dependence of the spectral moments alone.
This noncommutativity is the origin of nonlinear curvature flow.

The scale dependence of the curvature-conditioned point process is conveniently characterized by the dimensionless quantities
\begin{align}
	\gamma(R)
	=
	\frac{\sigma_1^2(R)}
	{\sigma_0(R)\sigma_2(R)},
	\label{eq:scale_dependent_gamma}
\end{align}
and
\begin{align}
	R_\ast(R)
	=
	\sqrt{3}
	\frac{\sigma_1(R)}
	{\sigma_2(R)}.
	\label{eq:scale_dependent_Rstar}
\end{align}
Here $\gamma(R)$ measures the coupling between amplitude and curvature at smoothing scale $R$, while $R_\ast(R)$ gives the characteristic geometric scale of peaks.

\begin{figure}[t]
	\centering
	\includegraphics[width=0.7\textwidth]{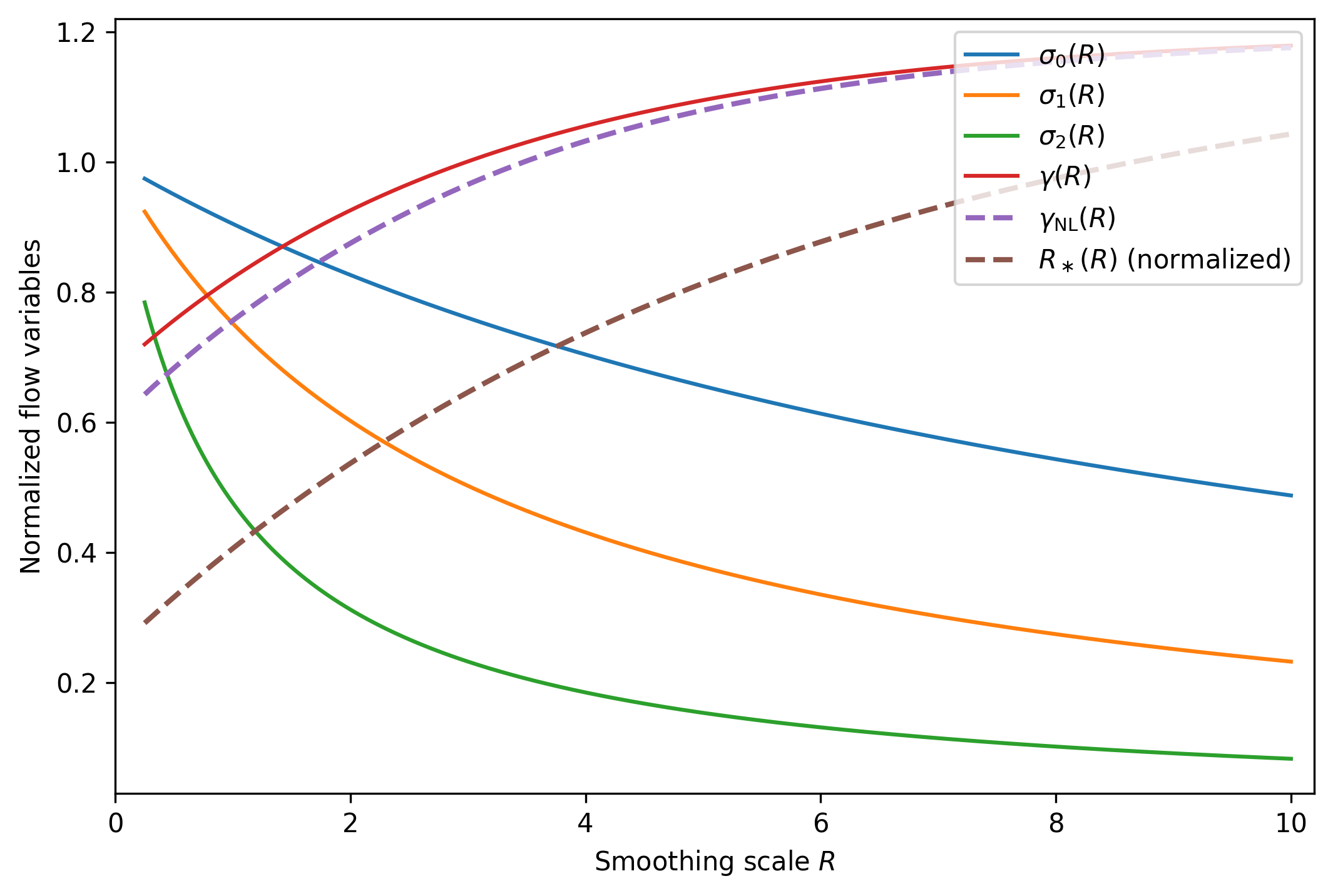}
	\caption{
	Illustration of the scale dependence of the curvature-flow variables.
	The figure shows the smoothing-scale dependence of the spectral moments $\sigma_j(R)$, the amplitude--curvature coupling parameter $\gamma(R)$ defined in Eq.~\eqref{eq:scale_dependent_gamma}, and the characteristic peak scale $R_\ast(R)$ defined in Eq.~\eqref{eq:scale_dependent_Rstar}, for a representative Gaussian smoothing flow.
	The dashed curve shows a representative nonlinear deformation of $\gamma(R)$ induced by the logarithmic entropy-response curvature discussed in Sections~\ref{sec:peaks} and \ref{subsec:coarse_grained_curvature}.
	This illustrates how curvature transport beyond the Gaussian closure deforms the amplitude--curvature coupling itself.
	}
	\label{fig:scale_flow}
\end{figure}

Figure~\ref{fig:scale_flow} illustrates the scale dependence of these quantities together with the spectral moments themselves for a representative smoothing flow.
The dashed curve additionally shows a representative nonlinear deformation of $\gamma(R)$ induced by the logarithmic entropy-response curvature.
The deviation from the Gaussian flow is largest at intermediate smoothing scales and gradually decreases toward large $R$, where coarse graining suppresses local nonlinear structure.
This behavior visualizes explicitly that nonlinear curvature transport modifies not only the peak abundance itself, but also the amplitude--curvature coupling governing the curvature-conditioned point process.

The important point is that, in the transport-geometric formulation, these quantities are not merely auxiliary parameters of a fixed-scale theory.
Rather, they represent explicit flow variables describing how the local curvature geometry changes under coarse graining.
The nonlinear shift of $\gamma(R)$ therefore provides a direct visualization of curvature transport beyond the Gaussian closure.

\subsection{Flow of the Logarithmic Curvature Tensor}
\label{subsec:flow_log_curvature}

The essential origin of nonlinear curvature flow is the noncommutativity between coarse graining and logarithmic transformation.
In general,
\begin{align}
	\ln(W_R\rho)
	\neq
	W_R(\ln\rho),
\end{align}
where $W_R$ denotes the smoothing operation at scale $R$.
Consequently, the scale evolution of the logarithmic curvature tensor cannot be reduced to a simple smoothing of the Gaussian-linear Hessian field.
The nonlinear flow therefore arises from the geometric response generated by this noncommutativity itself.

The scale flow of the curvature tensor is obtained by differentiating Eq.~\eqref{eq:scale_dependent_log_curvature} with respect to $R$.
We find
\begin{align}
	\partial_R H_{ij}(\bm{x};R)
	=
	-\partial_i\partial_j\partial_R\ln\rho_R(\bm{x}).
\end{align}
Since
\begin{align}
	\partial_R\ln\rho_R
	=
	\frac{\partial_R\rho_R}{\rho_R},
\end{align}
the curvature flow can be written as
\begin{align}
	\partial_R H_{ij}(\bm{x};R)
	=
	-\partial_i\partial_j
	\left(
		\frac{\partial_R\rho_R}{\rho_R}
	\right).
	\label{eq:curvature_flow_basic}
\end{align}
This expression is exact for a smooth positive coarse-grained density field.

Equation~\eqref{eq:curvature_flow_basic} is the basic flow equation for logarithmic curvature.
It shows that the change of curvature with smoothing scale is governed not by $\partial_R\rho_R$ alone, but by the relative scale response $\partial_R\rho_R/\rho_R$.
Thus, the flow is intrinsically multiplicative in density space and differs from the scale evolution of the ordinary Hessian of $\rho_R$.
Expanding Eq.~\eqref{eq:curvature_flow_basic}, one obtains
\begin{align}
	\partial_R H_{ij}
	&=
	-\frac{1}{\rho_R}\partial_i\partial_j\partial_R\rho_R
	+
	\frac{1}{\rho_R^2}
	\left[
		\partial_i\rho_R\,\partial_j\partial_R\rho_R
		+
		\partial_j\rho_R\,\partial_i\partial_R\rho_R
		+
		\partial_R\rho_R\,\partial_i\partial_j\rho_R
	\right] \notag \\
	&\quad-
	\frac{2\,\partial_R\rho_R}{\rho_R^3}
	\partial_i\rho_R\,\partial_j\rho_R .
	\label{eq:curvature_flow_expanded}
\end{align}
This form explicitly displays the nonlinear nature of the scale flow.
Even if $\partial_R\rho_R$ is determined by a linear smoothing operation, the induced flow of $H_{ij}$ contains nonlinear couplings between the density, its spatial derivatives, and its scale derivative.

In the linear Gaussian approximation, Eq.~\eqref{eq:curvature_flow_basic} reduces to
\begin{align}
	\partial_R H_{ij}
	\simeq
	-\partial_i\partial_j\partial_R\delta_R.
\end{align}
Therefore, the usual scale dependence of the BBKS variables is recovered as the linearized form of the logarithmic curvature flow.
The full equation, however, shows that nonlinear scale evolution transports curvature configurations in tensor space rather than merely changing their variance.

\subsection{Scale Flow of the Curvature-Conditioned Point Process}
\label{subsec:scale_flow_point_process}

At each smoothing scale, the peak point process is defined by the scale-dependent variables $(\rho_R,g_R,H_R)$.
The mean peak abundance is therefore
\begin{align}
	n_{\rm pk}(R)
	=
	\int
	\pd\rho_R\,\pd^3 g_R\,\pd H_R\,
	|\det H_R|\,
	p_R(\rho_R,g_R,H_R)\,
	\deltadir^{(3)}(g_R)\,
	\Theta(H_R).
	\label{eq:scale_dependent_peak_abundance}
\end{align}
Here $p_R$ denotes the joint distribution of the coarse-grained local variables.
Equation~\eqref{eq:scale_dependent_peak_abundance} is not a new peak-counting prescription.
It is the same curvature-conditioned point process evaluated on the scale-dependent tensor field.

Differentiating Eq.~\eqref{eq:scale_dependent_peak_abundance} with respect to $R$, we obtain
\begin{align}
	\partial_R n_{\rm pk}(R)
	=
	\int
	\pd\rho_R\,\pd^3 g_R\,\pd H_R\,
	\partial_R
	\left[
		|\det H_R|\,
		p_R(\rho_R,g_R,H_R)\,
		\deltadir^{(3)}(g_R)\,
		\Theta(H_R)
	\right].
	\label{eq:scale_flow_peak_abundance}
\end{align}
Equation~\eqref{eq:scale_flow_peak_abundance} shows that the scale flow of peak abundance has three distinct sources:
\begin{itemize}

\item
\textbf{Distributional flow.}
The joint distribution $p_R$ changes with scale.
In the linear Gaussian limit, this dependence is encoded entirely in the scale dependence of the spectral moments.
Beyond that limit, however, the flow includes higher-order cumulants and nonlinear couplings among local tensor variables.

\item
\textbf{Curvature-measure flow.}
The curvature tensor $H_R$ itself evolves according to Eq.~\eqref{eq:curvature_flow_basic}.
This changes the determinant factor $|\det H_R|$ and therefore directly deforms the induced point-process measure.

\item
\textbf{Selection-boundary flow.}
The positive-curvature condition is itself scale dependent because the tensor $H_R$ may enter or leave the cone $\mathcal{C}_+$ as $R$ changes.
This corresponds to the creation, destruction, or reclassification of peak configurations under coarse graining.

\end{itemize}
Thus, the scale dependence of peak statistics is decomposed into
\begin{align}
	\text{distributional flow}
	\quad+\quad
	\text{curvature-measure flow}
	\quad+\quad
	\text{selection-boundary flow}.
\end{align}
This decomposition is one of the main advantages of the curvature formulation.
It separates the change of the probability law from the change of the geometric measure and from the change of the admissible curvature domain.

\subsection{Eigenvalue Transport and the Positive Curvature Cone}
\label{subsec:eigenvalue_transport_cone}

The curvature flow can also be expressed in eigenvalue space.
Let
\begin{align}
	H_R e_a(R)=\lambda_a(R)e_a(R)
\end{align}
be the eigensystem of the scale-dependent curvature tensor.
For nondegenerate eigenvalues, the scale derivative of the $a$-th principal curvature is
\begin{align}
	\partial_R\lambda_a(R)
	=
	e_a^i(R)
	\left(\partial_R H_{ij}(R)\right)
	e_a^j(R).
	\label{eq:eigenvalue_scale_flow}
\end{align}
Substituting Eq.~\eqref{eq:curvature_flow_basic}, this becomes
\begin{align}
	\partial_R\lambda_a(R)
	=
	-
	e_a^i(R)
	\partial_i\partial_j
	\left(
		\frac{\partial_R\rho_R}{\rho_R}
	\right)
	e_a^j(R).
	\label{eq:eigenvalue_flow_log_density}
\end{align}
Thus, each principal curvature is transported in eigenvalue space by the second spatial derivative of the relative scale response of the coarse-grained density.

The positive-curvature condition at scale $R$ is
\begin{align}
	\lambda_a(R)>0
	\qquad
	(a=1,2,3).
\end{align}
A peak configuration remains a peak under an infinitesimal change of scale if all eigenvalues remain positive.
When one eigenvalue crosses zero, the tensor reaches the boundary of the positive curvature cone.
Therefore, the boundary condition
\begin{align}
	\lambda_a(R)=0
\end{align}
marks a scale-dependent transition of the local curvature type.
This provides a geometric interpretation of peak creation and destruction under smoothing.
A local structure may be selected as a peak at one scale and cease to be selected at another scale, not because the point-process formula changes, but because the curvature tensor is transported across the boundary of $\mathcal{C}_+$.
In this sense, the scale dependence of peak statistics is governed by motion in curvature tensor space.
The determinant factor is also transported along the flow.
For nondegenerate $H_R$,
\begin{align}
	\partial_R\ln|\det H_R|
	=
	\Tr\left(H_R^{-1}\partial_R H_R\right)
	=
	\sum_{a=1}^3
	\frac{\partial_R\lambda_a}{\lambda_a}.
	\label{eq:determinant_scale_flow}
\end{align}
This expression shows that the peak measure is particularly sensitive to scale evolution near the boundary of the positive curvature cone, where one or more $\lambda_a$ becomes small.
Thus, weakly curved structures are the most susceptible to reclassification under coarse graining.

\subsection{Gaussian Fixed-Scale Theory as a Solvable Section}
\label{subsec:gaussian_fixed_scale_section}

The scale-flow formulation clarifies the status of the conventional Gaussian peak theory.
In the BBKS limit, one fixes a smoothing scale $R$ and assumes that the local tensor variables follow a Gaussian distribution determined by the spectral moments $\sigma_j(R)$.
At that fixed scale, the curvature-conditioned point process admits a closed-form evaluation.
This corresponds to selecting a solvable section of the full scale-dependent theory.

More explicitly, the Gaussian fixed-scale theory is obtained by imposing
\begin{align}
	\begin{array}{c}
		\text{linear curvature}\\[3pt]
		+\\[3pt]
		\text{Gaussian closure}\\[3pt]
		+\\[3pt]
		\text{fixed smoothing scale}.
	\end{array}
\end{align}
Under these conditions, the scale dependence enters only through the spectral moments and their dimensionless combinations.
The curvature tensor itself is not treated as flowing in a nonlinear tensor space.
The positive-curvature cone is sampled at a fixed value of $R$, and the point-process measure is evaluated at that scale.

In the present formulation, this is only one slice of a larger structure.
Changing $R$ transports the curvature tensor, changes the distribution of local variables, and moves configurations relative to the positive curvature cone.
Therefore, BBKS should be understood as a fixed-scale Gaussian section of the curvature-flow theory, not as the full structure of peak statistics.

This distinction is important for applications to simulations and observations.
In practical analyses, peaks identified in smoothed density fields depend sensitively on the smoothing scale.
The present formulation provides a way to describe this dependence geometrically, through the flow of curvature tensors and the induced flow of the peak measure.
Thus, scale dependence becomes part of the theory rather than an external choice made after the theory has been formulated.

\section{Response Hierarchy of Curvature-Conditioned Peaks}
\label{sec:response_hierarchy}

We now turn to the observable consequences of these structures.
The central viewpoint of this section is that \emph{peak statistics are naturally organized as a hierarchy of response functions to long-wavelength background fields.}

In the conventional Gaussian peak theory, the primary quantities are the peak abundance, the two-point correlation, and the associated large-scale bias.
In the present framework, however, these quantities are understood as different levels of a unified response hierarchy generated by the curvature-conditioned point process.
The essential point is that the local geometric constraints defining peaks remain fixed, while the background field modulates the joint distribution and the curvature measure associated with these constraints.

This perspective extends the idea of response-based bias theories into the geometry of local compression structures.
The response of peak statistics is no longer interpreted merely as the modulation of abundance thresholds.
Instead, it measures how the geometry of the curvature tensor field itself reacts to large-scale background deformations.
In this sense, the theory developed here is not only a theory of peak counting, but also a theory of geometric susceptibilities of curvature-selected structures.

The hierarchy constructed below proceeds systematically from the one-point response to the response of higher-order peak correlations.
As will become clear, the conventional peak bias corresponds only to the first member of this hierarchy.

\subsection{Linear Response of Curvature-Conditioned Peaks}
\label{subsec:linear_response_peaks}

We first consider the response of the one-point peak statistic to a long-wavelength background fluctuation.
Let $\delta_{\rm L}(\bm{x})$ denote a coarse-grained large-scale mode defined by
\begin{align}
	\delta_{\rm L}(\bm{x})
	=
	\int
	\pd^3 y\,
	W_{\rm L}(|\bm{x}-\bm{y}|)
	\delta(\bm{y}),
\end{align}
where $W_{\rm L}$ is a smoothing kernel with characteristic scale $R_{\rm L}$ much larger than the local peak scale.
The field $\delta_{\rm L}$ therefore acts as a slowly varying background deformation of the local density environment.

The peak number density depends on the local joint distribution of amplitude, gradient, and curvature variables.
Consequently, the background field changes not only the local density threshold, but also the curvature distribution and the induced geometric measure.
The response of the peak abundance is therefore defined by
\begin{align}
	R_{\rm pk}^{(1)}
	\equiv
	\frac{\partial\ln n_{\rm pk}}{\partial\delta_{\rm L}}.
	\label{eq:first_response_definition}
\end{align}
This quantity is the first response coefficient of the curvature-conditioned point process.
In the conventional terminology of peak theory, it corresponds to the large-scale peak bias,
\begin{align}
	b_{\rm pk}
	=
	R_{\rm pk}^{(1)}.
\end{align}

The important point, however, is that Eq.~\eqref{eq:first_response_definition} is not interpreted here merely as the response of a thresholded abundance.
The peak number density is defined by the curvature-conditioned point-process measure,
\begin{align}
	n_{\rm pk}
	=
	\left\langle
		|\det H|\,
		\deltadir^{(3)}(g)\,
		\Theta(H)
	\right\rangle.
\end{align}
Therefore, the background field affects three structures simultaneously:
\begin{itemize}

\item
the local amplitude distribution,

\item
the statistics of the curvature tensor,

\item
the induced geometric measure through the determinant factor.

\end{itemize}
Thus, the linear response probes not merely the abundance of peaks, but the susceptibility of the local curvature geometry to long-wavelength deformation.

In the Gaussian-linear limit, the response coefficient reduces to
\begin{align}
	R_{\rm pk}^{(1)}
	=
	\frac{1}{\sigma_0}
	\left[
		\nu
		-
		\gamma
		\langle x\rangle_{\rm pk}
	\right],
	\label{eq:first_response_gaussian}
\end{align}
where $\nu$ is the peak height, $x$ is the dimensionless curvature invariant, and $\langle x\rangle_{\rm pk}$ denotes the curvature averaged over the peak ensemble.
Equation~\eqref{eq:first_response_gaussian} already shows that the response naturally decomposes into two qualitatively distinct contributions:
\begin{align}
	R_{\rm pk}^{(1)}
	=
	R_{\rm amp}^{(1)}
	+
	R_{\rm curv}^{(1)}.
\end{align}
The first term represents the amplitude response associated with density-threshold modulation, while the second term represents the curvature response associated with deformation of the local compression geometry.

This decomposition is conceptually important.
In conventional local-threshold theories, bias is primarily associated with amplitude selection.
In the present formulation, however, the response contains an intrinsically geometric contribution originating from the curvature tensor field.
Therefore, the peak bias is interpreted not simply as an abundance response, but as the first geometric susceptibility of the curvature-conditioned point process.

This viewpoint also clarifies why peak statistics are naturally connected to higher-order response theory.
Since the point-process operator itself depends on the curvature tensor, its response to a background field automatically propagates into higher-order joint statistics.
The one-point response is therefore only the lowest level of a larger hierarchy describing the modulation of local compression structures by long-wavelength environments.

\subsection{Two-Point Response Functions and Joint Curvature Structure}
\label{subsec:two_point_response}

We next consider the response of the two-point peak correlation to a long-wavelength background field.
The essential point is that, once peaks are defined as curvature-selected structures, the response of the two-point statistic measures not merely the modulation of pair abundance, but the deformation of the joint geometry of local compression centers.

The two-point peak function is defined by
\begin{align}
	\rho_{\rm pk}^{(2)}(\bm{x}_1,\bm{x}_2)
	=
	\left\langle
		n_{\rm pk}(\bm{x}_1)\,
		n_{\rm pk}(\bm{x}_2)
	\right\rangle.
\end{align}
Assuming statistical homogeneity and isotropy, this depends only on the separation
\begin{align}
	r=|\bm{x}_2-\bm{x}_1|,
\end{align}
and the corresponding response coefficient is defined as
\begin{align}
	R_{\rm pk}^{(2)}(r)
	\equiv
	\frac{\partial\ln\rho_{\rm pk}^{(2)}(r)}
	{\partial\delta_L}.
	\label{eq:two_point_response_definition}
\end{align}
Equivalently, in terms of the usual two-point correlation function,
\begin{align}
	1+\xi_{\rm pk}(r)
	=
	\frac{\rho_{\rm pk}^{(2)}(r)}
	{\bar n_{\rm pk}^2},
\end{align}
one may define
\begin{align}
	R_{\xi}(r)
	=
	\frac{\partial\ln[1+\xi_{\rm pk}(r)]}
	{\partial\delta_L}.
\end{align}

Unlike the one-point response, the two-point response contains information about the deformation of the joint curvature configuration between two positions.
This is because the underlying two-point peak function is determined not only by the local peak conditions at each point, but also by the joint distribution of the local tensor variables.
In the notation introduced previously,
\begin{align}
	\Gamma_a=(\rho_a,g_a,H_a),
	\qquad
	a=1,2,
\end{align}
the two-point peak function is given by
\begin{align}
	\rho_{\rm pk}^{(2)}(r)
	=
	\int
	\pd\Gamma_1\,\pd\Gamma_2\,
	|\det H_1|\,|\det H_2|\,
	p_2(\Gamma_1,\Gamma_2;r)\,
	\deltadir^{(3)}(g_1)\,
	\deltadir^{(3)}(g_2)\,
	\Theta(H_1)\,
	\Theta(H_2).
\end{align}
Therefore, the background field modulates not only the local abundance at each point, but also the joint distribution $p_2$ itself.
The two-point response is thus sensitive to the susceptibility of the pairwise curvature structure.

In the Gaussian-linear limit, the joint distribution is completely determined by the correlation function $\xi(r)$ and its derivatives.
In particular, the curvature-tensor correlation is
\begin{align}
	\langle
	H_{ij}(\bm{x})
	H_{kl}(\bm{x}+\bm{r})
	\rangle
	=
	\partial_i\partial_j\partial_k\partial_l
	\xi(r).
	\label{eq:curvature_tensor_2pt_response}
\end{align}
This relation is central to the geometric interpretation of the response hierarchy.
The response of the two-point peak correlation is controlled not merely by the amplitude correlation $\xi(r)$ itself, but by the background modulation of the spatial correlation of local compression structures.

Differentiating with respect to the long-wavelength background field, the leading geometric response kernels are therefore determined by
\begin{align}
	\frac{\partial\xi(r)}{\partial\delta_L},
	\qquad
	\frac{\partial}{\partial\delta_L}
	\partial_i\partial_j\xi(r),
	\qquad
	\frac{\partial}{\partial\delta_L}
	\partial_i\partial_j\partial_k\partial_l\xi(r).
\end{align}
The first term represents the conventional modulation of the density correlation.
The second term describes the response of the gradient structure.
The third term is the response of the joint curvature geometry itself.

This decomposition clarifies an important distinction.
The ordinary density two-point correlation probes how strongly density amplitudes are correlated between two points.
The peak two-point response, by contrast, probes how strongly the local compression geometry reacts coherently to a background deformation.
In this sense, the response function measures the susceptibility of the transport-induced curvature structure rather than merely the susceptibility of density amplitudes.

The two-point response therefore contains a genuinely geometric contribution absent from ordinary threshold models.
Even in the Gaussian limit, the background field modulates not only the abundance of peaks but also the correlation of their curvature tensors.
Consequently, the two-point response should be interpreted as the first nontrivial susceptibility of the joint curvature geometry.

This viewpoint also clarifies the relation to the nonlinear curvature transport discussed in the previous sections.
As the curvature tensor evolves under nonlinear deformation and scale flow, the joint curvature correlations evolve simultaneously.
The response hierarchy therefore organizes not only the statistics of peaks themselves, but also the propagation of geometric information across spatial scales and between multiple local compression centers.

\subsection{Three-Point Response and Joint Curvature Geometry}
\label{subsec:three_point_response}

We now extend the response formulation to the three-point statistic.
While the two-point response probes the modulation of pairwise compression structures, the three-point response measures how the geometry of triangular configurations of local compression centers reacts to a long-wavelength background field.
In this sense, it provides \emph{the first genuinely higher-order geometric susceptibility of the curvature-conditioned point process.}

The three-point peak function is defined by
\begin{align}
	\rho_{\rm pk}^{(3)}
	(\bm{x}_1,\bm{x}_2,\bm{x}_3)
	=
	\left\langle
		n_{\rm pk}(\bm{x}_1)\,
		n_{\rm pk}(\bm{x}_2)\,
		n_{\rm pk}(\bm{x}_3)
	\right\rangle.
\end{align}
Under statistical homogeneity and isotropy, this quantity depends only on the triangle configuration specified by
\begin{align}
	r_{12}=|\bm{x}_1-\bm{x}_2|,
	\qquad
	r_{23}=|\bm{x}_2-\bm{x}_3|,
	\qquad
	r_{31}=|\bm{x}_3-\bm{x}_1|.
\end{align}
The associated response coefficient is therefore defined as
\begin{align}
	R_{\rm pk}^{(3)}
	(r_{12},r_{23},r_{31})
	\equiv
	\frac{
		\partial
		\ln
		\rho_{\rm pk}^{(3)}
		(r_{12},r_{23},r_{31})
	}{
		\partial\delta_L
	}.
	\label{eq:three_point_response_definition}
\end{align}

The important point is that the three-point response is sensitive not merely to three-body abundance correlations, but to the compatibility of local curvature geometry across a triangular configuration.
This follows from the structure of the underlying point-process operator,
\begin{align}
	n_{\rm pk}(\bm{x})
	=
	|\det H(\bm{x})|\,
	\deltadir^{(3)}(g(\bm{x}))\,
	\Theta(H(\bm{x})).
\end{align}
The three-point statistic therefore contains three simultaneous curvature constraints together with the joint distribution of the local tensor variables at the three positions.

Introducing
\begin{align}
	\Gamma_a=(\rho_a,g_a,H_a),
	\qquad
	a=1,2,3,
\end{align}
the three-point function is written as
\begin{align}
	\rho_{\rm pk}^{(3)}
	=
	\int
	\prod_{a=1}^3
	\left[
		\pd\Gamma_a\,
		|\det H_a|\,
		\deltadir^{(3)}(g_a)\,
		\Theta(H_a)
	\right]
	p_3(\Gamma_1,\Gamma_2,\Gamma_3).
	\label{eq:three_point_response_integral}
\end{align}
Thus, the three-point response probes the deformation of the three-point joint distribution $p_3$ under the background field.

An important feature appears already in the Gaussian limit.
Even if the underlying density field is Gaussian, the connected three-point peak correlation does not generally vanish,
\begin{align}
	\zeta_{\rm pk}^{(\mathrm{G})}\neq 0,
\end{align}
because the peak operator itself is nonlinear.
This fact is conceptually important.
The geometric nonlinearity of the curvature-conditioned point process produces irreducible higher-order correlations independently of primordial non-Gaussianity.

From the response viewpoint, this means that the three-point susceptibility naturally decomposes into two qualitatively distinct parts.
The first originates from the nonlinear conditioning imposed by the curvature-selected point process itself.
The second originates from the deformation of the underlying distribution.
Symbolically,
\begin{align}
	R_{\rm pk}^{(3)}
	=
	R_{\rm geom}^{(3)}
	+
	\Delta R_{\rm dist}^{(3)}.
	\label{eq:three_point_response_decomposition}
\end{align}
The first term is present even in the Gaussian reference theory and reflects the geometric structure of local compression centers.
The second term encodes the additional response induced by non-Gaussian deformation of the joint distribution.

This decomposition provides a clear interpretation of higher-order peak statistics.
The three-point response is not merely a diagnostic of primordial non-Gaussianity.
It is first of all a geometric susceptibility measuring how triangular configurations of local compression structures react collectively to a large-scale deformation.
Primordial non-Gaussianity and nonlinear transport appear as additional deformations of this underlying geometric response.

The geometrical meaning becomes particularly transparent from the transport viewpoint introduced earlier.
Since the curvature tensor measures the local variation of the transport Jacobian, the three-point response describes how the mutual arrangement of three local compression centers changes under a background deformation of the transport flow.
Thus, while the two-point response probes correlated compression, the three-point response probes correlated geometric organization of compression structures.

This distinction is essential for the hierarchy developed below.
The one-point response measures the susceptibility of local abundance.
The two-point response measures the susceptibility of pairwise compression structure.
The three-point response measures the susceptibility of triangular geometric organization.
The hierarchy therefore systematically increases the level of geometric complexity probed by the curvature-conditioned point process.

\subsection{General Response Hierarchy}
\label{subsec:general_response_hierarchy}

The structures derived above suggest a natural unification of peak statistics.
The one-point response, the two-point response, and the three-point response are not independent quantities.
They are successive levels of a single response hierarchy generated by the curvature-conditioned point process.

For the general $n$-point peak function,
\begin{align}
	\rho_{\rm pk}^{(n)}
	(\bm{x}_1,\dots,\bm{x}_n)
	=
	\left\langle
		\prod_{a=1}^n
		n_{\rm pk}(\bm{x}_a)
	\right\rangle,
\end{align}
the corresponding response coefficient is defined by
\begin{align}
	R_{\rm pk}^{(n)}
	(\bm{x}_1,\dots,\bm{x}_n)
	\equiv
	\frac{
		\partial
		\ln
		\rho_{\rm pk}^{(n)}
		(\bm{x}_1,\dots,\bm{x}_n)
	}{
		\partial\delta_L
	}.
	\label{eq:general_response_hierarchy_definition}
\end{align}
Thus, the full hierarchy of peak statistics may be organized as
\begin{align}
	\left\{
		n_{\rm pk},
		\rho_{\rm pk}^{(2)},
		\rho_{\rm pk}^{(3)},
		\dots
	\right\}
	\quad\longleftrightarrow\quad
	\left\{
		R_{\rm pk}^{(1)},
		R_{\rm pk}^{(2)},
		R_{\rm pk}^{(3)},
		\dots
	\right\},
\end{align}
with
\begin{align}
	R_{\rm pk}^{(1)}
	=
	b_{\rm pk}.
\end{align}

This hierarchy provides a unifying interpretation of peak statistics.
The individual correlation functions are reinterpreted as geometric susceptibilities of the curvature-selected point process under long-wavelength deformation.
The order of the statistic determines the level of collective geometric organization being probed.

More specifically:
\begin{itemize}

\item
The one-point response measures the susceptibility of local peak abundance.

\item
The two-point response measures the susceptibility of pairwise compression structure.

\item
The three-point response measures the susceptibility of triangular geometric organization.

\item
Higher-order responses measure increasingly complex collective arrangements of local curvature structures.

\end{itemize}

An important property of this hierarchy is that the local geometric structure remains unchanged at every order.
The zero-gradient condition, the positive-curvature constraint, and the determinant Jacobian continue to appear identically at each point.
The increase in complexity arises entirely from the extension of the joint distribution and from the higher-order response of the associated curvature geometry.

This separation between fixed geometric structure and extended joint response is one of the central organizational principles of the present theory.
The hierarchy is therefore not generated by introducing new geometric ingredients at each order.
Instead, the same local curvature-selection rule is repeated while the collective structure of the distribution becomes increasingly rich.

The response hierarchy also clarifies the role of the Gaussian-linear theory.
In that limit, the hierarchy is determined by derivatives of finite-dimensional Gaussian measures.
For example, the two-point response is controlled by the background dependence of
\begin{align}
	\xi(r),
	\qquad
	\partial_i\partial_j\xi(r),
	\qquad
	\partial_i\partial_j\partial_k\partial_l\xi(r),
\end{align}
while higher-order responses involve the corresponding higher-order joint kernels.
Thus, the Gaussian theory corresponds to the analytically solvable sector of the hierarchy.
Beyond Gaussianity, however, the hierarchy acquires additional response channels associated with higher-order cumulants of the joint distribution.
Importantly, these corrections modify the distributional sector without changing the underlying curvature-conditioned structure.
In this sense, primordial non-Gaussianity, nonlinear transport, scale flow, and weighted measures all enter the hierarchy as deformations of the response kernels rather than as replacements of the geometric framework itself.

The hierarchy constructed here therefore reorganizes peak statistics into a theory of geometric response functions.
From this viewpoint, the classical BBKS quantities correspond only to the first few levels of a much larger transport-geometric structure.

\subsection{Response Decomposition and Geometric Susceptibility}
\label{subsec:response_decomposition}

The response hierarchy introduced above becomes operationally useful once the response coefficients are decomposed according to the structures entering the curvature-conditioned point process.
Since
\begin{align}
	n_{\rm pk}(\bm{x})
	=
	|\det H(\bm{x})|\,
	\deltadir^{(3)}(g(\bm{x}))\,
	\Theta(H(\bm{x})),
\end{align}
the response to a long-wavelength background field $\delta_L$ acts simultaneously on the joint distribution, the curvature measure, and the peak-selection boundary.

Let the scale-dependent joint distribution of local variables be written schematically as
\begin{align}
	p(\Gamma|\delta_L),
	\qquad
	\Gamma=(\rho,g,H).
\end{align}
Then the one-point response coefficient is
\begin{align}
	R_{\rm pk}^{(1)}
	=
	\frac{\partial\ln n_{\rm pk}}{\partial\delta_L}
	=
	\frac{1}{n_{\rm pk}}
	\frac{\partial n_{\rm pk}}{\partial\delta_L},
\end{align}
with
\begin{align}
	\frac{\partial n_{\rm pk}}{\partial\delta_L}
	=
	\int
	\pd\Gamma\,
	\partial_{\delta_L}
	\left[
		|\det H|\,
		p(\Gamma|\delta_L)\,
		\deltadir^{(3)}(g)\,
		\Theta(H)
	\right].
\end{align}
Expanding the derivative gives
\begin{align}
	\frac{\partial n_{\rm pk}}{\partial\delta_L}
	=
	I_{\rm dist}
	+
	I_{\rm det}
	+
	I_{\Theta},
	\label{eq:response_three_sector_split}
\end{align}
where
\begin{align}
	I_{\rm dist}
	&=
	\int
	\pd\Gamma\,
	|\det H|\,
	(\partial_{\delta_L}p)\,
	\deltadir^{(3)}(g)\,
	\Theta(H),
\end{align}
\begin{align}
	I_{\rm det}
	&=
	\int
	\pd\Gamma\,
	(\partial_{\delta_L}|\det H|)\,
	p\,
	\deltadir^{(3)}(g)\,
	\Theta(H),
\end{align}
and
\begin{align}
	I_{\Theta}
	&=
	\int
	\pd\Gamma\,
	|\det H|\,p\,
	\deltadir^{(3)}(g)\,
	(\partial_{\delta_L}\Theta(H)).
\end{align}

Equation~\eqref{eq:response_three_sector_split} provides the basic decomposition of the response hierarchy.
The first term describes deformation of the joint distribution.
The second term describes deformation of the curvature measure itself.
The third term describes motion of the curvature tensor relative to the boundary of the positive curvature cone.

The determinant sector can be evaluated explicitly using the nonlinear curvature expansion developed in Section~\ref{sec:nonlinear_deformation}.
Writing
\begin{align}
	H
	=
	H^{(0)}
	+
	H^{(2)},
\end{align}
the determinant response is
\begin{align}
	\partial_{\delta_L}\ln|\det H|
	=
	\Tr(H^{-1}\partial_{\delta_L}H).
\end{align}
To leading order,
\begin{align}
	\partial_{\delta_L}\ln|\det H|
	\simeq
	\sum_{a=1}^3
	\frac{
		e_a^i
		(\partial_{\delta_L}H_{ij})
		e_a^j
	}{
		\lambda_a
	},
	\label{eq:det_response_eigen}
\end{align}
where $\lambda_a$ and $e_a$ are the principal curvatures and eigenvectors of $H$.
Thus, the response of the geometric measure is enhanced near weakly curved directions where $\lambda_a$ is small.

The two-point response function is obtained similarly.
Defining
\begin{align}
	R_{\rm pk}^{(2)}(r)
	=
	\frac{
		\partial
		\ln\rho_{\rm pk}^{(2)}(r)
	}{
		\partial\delta_L
	},
\end{align}
we obtain
\begin{align}
	\partial_{\delta_L}
	\rho_{\rm pk}^{(2)}
	=
	I_{\rm dist}^{(2)}
	+
	I_{\rm det}^{(2)}
	+
	I_{\Theta}^{(2)},
\end{align}
with
\begin{align}
	I_{\rm dist}^{(2)}
	&=
	\int
	\pd\Gamma_1\pd\Gamma_2\,
	|\det H_1||\det H_2|\,
	(\partial_{\delta_L}p_2)\,
	\prod_{a=1}^2
	\left[
		\deltadir^{(3)}(g_a)\Theta(H_a)
	\right],
\end{align}
\begin{align}
	I_{\rm det}^{(2)}
	&=
	\int
	\pd\Gamma_1\pd\Gamma_2\,
	\partial_{\delta_L}
	(|\det H_1||\det H_2|)
	\,p_2\,
	\prod_{a=1}^2
	\left[
		\deltadir^{(3)}(g_a)\Theta(H_a)
	\right].
\end{align}

In the Gaussian-linear limit, the distributional sector is determined by derivatives of the covariance kernels,
\begin{align}
	\partial_{\delta_L}\xi(r),
	\qquad
	\partial_{\delta_L}
	\partial_i\partial_j\xi(r),
	\qquad
	\partial_{\delta_L}
	\partial_i\partial_j
	\partial_k\partial_l
	\xi(r).
\end{align}
The last quantity is particularly important because it controls the response of the correlated curvature tensor itself:
\begin{align}
	\partial_{\delta_L}
	\langle
	H_{ij}(\bm{x})
	H_{kl}(\bm{x}+\bm{r})
	\rangle.
\end{align}
Thus, the two-point response probes how the background field modulates the correlation of local compression geometry rather than merely the correlation of density amplitudes.

The same structure extends to arbitrary order.
For the $n$-point function,
\begin{align}
	\rho_{\rm pk}^{(n)}
	=
	\left\langle
		\prod_{a=1}^n
		n_{\rm pk}(\bm{x}_a)
	\right\rangle,
\end{align}
the response hierarchy is generated recursively by
\begin{align}
	\partial_{\delta_L}
	\rho_{\rm pk}^{(n)}
	=
	I_{\rm dist}^{(n)}
	+
	I_{\rm det}^{(n)}
	+
	I_{\Theta}^{(n)}.
	\label{eq:npoint_response_recursive}
\end{align}
Therefore, the hierarchy is not merely an increase in the order of correlation functions.
It is a hierarchy of coupled responses of \emph{distribution}, \emph{curvature measure}, and \emph{selection geometry}.
This decomposition is one of the main structural consequences of the present framework.
The Gaussian BBKS theory corresponds to the special limit in which the distributional sector is Gaussian, the curvature response is linearized, and the cone response is fixed at a single smoothing scale.
Beyond that limit, the response hierarchy systematically organizes how nonlinear transport, non-Gaussianity, and scale flow deform the geometry of local compression structures.

\begin{figure}[t]
	\centering
	\includegraphics[width=0.7\textwidth]{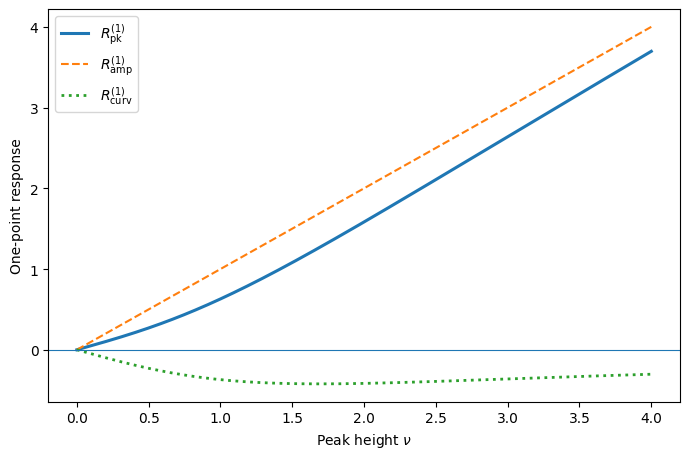}
    \includegraphics[width=0.7\textwidth]{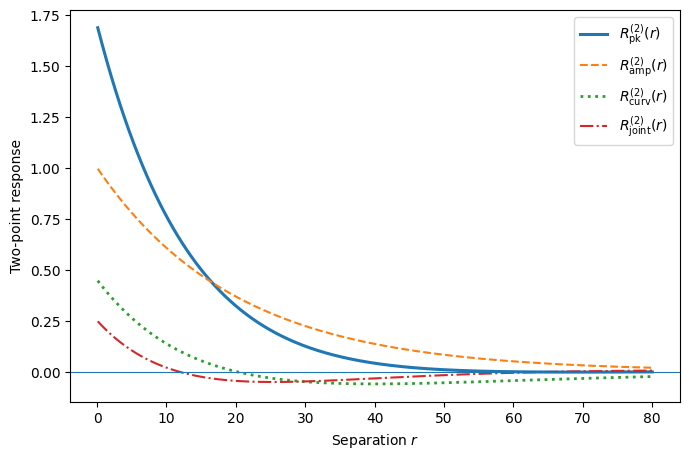}
    \caption{
	Illustration of the response hierarchy of the curvature-conditioned point process.
	The upper-level response is decomposed into amplitude, curvature, and joint-geometry sectors as defined in Eqs.~\eqref{eq:illustrative_one_point_response} and \eqref{eq:illustrative_two_point_response}.
	The left panel shows the representative one-point response as a function of peak height $\nu$, while the right panel shows the representative two-point response as a function of separation $r$.
	The joint-geometry sector appears first at the two-point level and describes the response of correlated local compression structures.
	}
	\label{fig:response_hierarchy}
\end{figure}

For illustration, we decompose the response functions into amplitude, curvature, and joint-geometry sectors as
\begin{align}
	R_{\rm pk}^{(1)}(\nu)
	&=
	R_{\rm amp}^{(1)}(\nu)
	+
	R_{\rm curv}^{(1)}(\nu),
	\label{eq:illustrative_one_point_response}
\end{align}
and
\begin{align}
	R_{\rm pk}^{(2)}(r)
	&=
	R_{\rm amp}^{(2)}(r)
	+
	R_{\rm curv}^{(2)}(r)
	+
	R_{\rm joint}^{(2)}(r).
	\label{eq:illustrative_two_point_response}
\end{align}
Representative analytic forms are plotted in Fig.~\ref{fig:response_hierarchy}.

At the level of perturbative large-scale bias expansions, the response coefficients introduced here play roles analogous to conventional Lagrangian and Eulerian bias parameters.
For example, the amplitude-response sector corresponds to density-type bias coefficients, while the curvature and joint-geometry sectors naturally generate tensorial and tidal-type responses associated with operators such as
$s_{ij}s^{ij}$.
In this sense, the present response hierarchy provides a geometric organization of higher-order bias structures in terms of curvature-conditioned measures.

\section{Beyond Gaussian Closure and Weighted Curvature Geometry}
\label{sec:beyond_gaussian}

The response hierarchy developed in the previous section was formulated without assuming that the underlying distribution must remain Gaussian.
The Gaussian-linear theory provides a solvable reference structure, but the curvature-conditioned point process itself is more general.
The purpose of this section is to clarify how the hierarchy is deformed once Gaussian closure is relaxed and once weighted geometric measures are introduced.

The important point is that the underlying geometric skeleton remains unchanged.
The zero-gradient condition, the positive-curvature constraint, and the determinant Jacobian continue to define the local peak-selection structure.
What changes beyond the Gaussian limit is the distributional sector and the weighting of the induced geometric measure.
In this sense, non-Gaussianity and weighted curvature geometry should be understood not as replacements of the framework, but as controlled deformations of the response hierarchy itself.

\subsection{Persistence of the Curvature-Conditioned Structure}
\label{subsec:persistence_structure}

The nonlinear deformation theory and the response hierarchy developed in the previous sections relied on the decomposition
\begin{align}
	\partial_{\delta_L}\rho_{\rm pk}^{(n)}
	=
	I_{\rm dist}^{(n)}
	+
	I_{\rm det}^{(n)}
	+
	I_{\Theta}^{(n)}.
\end{align}
This decomposition continues to hold even when Gaussianity is relaxed.
The reason is that it follows directly from the geometric structure of the curvature-conditioned point process rather than from the explicit form of the distribution.

Indeed, the local definition of peaks itself remains unchanged.
At every order of the hierarchy, peaks are selected by the vanishing of the gradient field together with the requirement that the curvature tensor belongs to the positive curvature cone.
The associated point-process measure is then induced by the determinant Jacobian of the curvature tensor.
These ingredients are independent of whether the underlying field is Gaussian or non-Gaussian, linear or nonlinear.
What changes beyond Gaussian closure is therefore not the geometric skeleton of the theory, but the joint distribution of the local tensor variables and the resulting response kernels.

This distinction is important conceptually.
The Gaussian theory does not define the geometry of peaks.
It only provides a finite-dimensional closed form for the associated joint distribution.
The curvature-conditioned structure itself is fixed before the distribution is specified.
From this viewpoint, the Gaussian-linear theory should be understood as a solvable sector embedded within a more general geometric response theory.

The same conclusion applies at every level of the hierarchy.
For the general response function,
\begin{align}
	R_{\rm pk}^{(n)}
	=
	R_{\rm amp}^{(n)}
	+
	R_{\rm curv}^{(n)}
	+
	R_{\rm joint}^{(n)},
\end{align}
the decomposition into amplitude, curvature, and joint-geometry sectors remains meaningful independently of Gaussianity.
Beyond Gaussian closure, the explicit form of these sectors is deformed by higher-order cumulants and nonlinear transport effects, but the decomposition itself persists.

In the Gaussian-linear limit, the hierarchy closes because the joint distribution is determined entirely by finite-dimensional covariance kernels.
Once this assumption is relaxed, new response channels appear through higher-order correlations among the local tensor variables.
Importantly, however, these extensions deform the response kernels rather than replacing the underlying curvature-conditioned structure itself.

The persistence of this geometric skeleton is one of the main structural consequences of the present framework.
It allows nonlinear transport, non-Gaussianity, scale flow, and weighted measures to be incorporated within a unified response-theoretic organization while preserving the same local curvature-selection rule throughout the hierarchy.

\subsection{Non-Gaussian Deformation of the Joint Distribution}
\label{subsec:nonGaussian_deformation}

We next consider the deformation of the response hierarchy once the joint distribution of local tensor variables deviates from Gaussianity.
The essential point is that non-Gaussianity modifies the distributional sector of the hierarchy while leaving the curvature-conditioned geometric structure intact.

Let
\begin{align}
	p_G(\Gamma)
\end{align}
denote the Gaussian reference distribution of the local variables
\begin{align}
	\Gamma=(\rho,g,H).
\end{align}
Beyond Gaussian closure, the full distribution may be written schematically as
\begin{align}
	p(\Gamma)
	=
	p_G(\Gamma)
	+
	\Delta p_{\rm NG}(\Gamma),
\end{align}
where $\Delta p_{\rm NG}$ contains the higher-order cumulant structure.
Under weak non-Gaussianity, this deformation can be organized systematically through an Edgeworth- or Gram--Charlier-type expansion.
For example, using the dimensionless amplitude $\nu$ and curvature invariant $x$, the distribution may be written as
\begin{align}
	p(\nu,x,\cdots)
	=
	p_G(\nu,x,\cdots)
	\left[
		1
		+
		\frac16\kappa_{300}H_3(\nu)
		+
		\frac12\kappa_{210}H_2(\nu)H_1(x)
		+
		\frac12\kappa_{120}H_1(\nu)H_2(x)
		+
		\cdots
	\right],
	\label{eq:edgeworth_curvature}
\end{align}
where $\kappa_{abc}$ are mixed cumulants and $H_n$ are Hermite polynomials.

Equation~\eqref{eq:edgeworth_curvature} shows that non-Gaussianity does not merely distort the one-point amplitude distribution.
It deforms the coupling between amplitude and curvature variables and therefore modifies the geometric response kernels themselves.
In particular, while the Gaussian theory is characterized by the single coupling parameter
\begin{align}
	\gamma
	=
	\frac{\sigma_1^2}{\sigma_0\sigma_2},
\end{align}
the non-Gaussian theory introduces an infinite hierarchy of independent mixed cumulants coupling the amplitude, gradient, and curvature sectors.

This deformation propagates directly into the response hierarchy.
For the general response coefficient,
\begin{align}
	R_{\rm pk}^{(n)}
	=
	R_{{\rm pk},G}^{(n)}
	+
	\Delta R_{\rm NG}^{(n)},
\end{align}
the non-Gaussian correction arises from the deformation of the distributional sector,
\begin{align}
	\Delta R_{\rm NG}^{(n)}
	\sim
	\partial_{\delta_L}
	\Delta p_{\rm NG}.
\end{align}
At leading order, the correction to the one-point response becomes
\begin{align}
	\Delta R_{\rm NG}^{(1)}
	\propto
	\kappa_{300}H_2(\nu)
	+
	\kappa_{210}H_1(\nu)H_1(x)
	+
	\cdots ,
\end{align}
showing explicitly that the background response acquires mixed amplitude--curvature contributions absent in the Gaussian theory.

The same structure appears at higher order.
For the two-point response,
\begin{align}
	R_{\rm pk}^{(2)}
	=
	R_{{\rm pk},G}^{(2)}
	+
	\Delta R_{\rm NG}^{(2)},
\end{align}
the correction term contains non-Gaussian deformations of the joint curvature kernels,
\begin{align}
	\Delta R_{\rm NG}^{(2)}
	\sim
	\partial_{\delta_L}
	\left[
		\langle HH\rangle_c,
		\langle \delta HH\rangle_c,
		\langle HHH\rangle_c,
		\dots
	\right],
\end{align}
where the connected cumulants involve the curvature tensor itself.
Thus, the response hierarchy is deformed not only through density-amplitude correlations, but through higher-order correlations of local compression geometry.

This distinction is physically important.
Primordial non-Gaussianity and nonlinear gravitational evolution both appear as deformations of the joint distribution.
The former enters through the initial higher-order cumulants, while the latter generates effective cumulants dynamically through nonlinear transport.
In both cases, however, the curvature-conditioned point-process structure remains unchanged.
The non-Gaussian effects act by deforming the response kernels rather than replacing the geometric framework.

From this viewpoint, non-Gaussianity should not be interpreted merely as a deviation from a Gaussian probability law.
It is more fundamentally a deformation of the geometry of the response hierarchy itself.
The Gaussian theory provides the analytically closed reference structure, while the higher-order cumulants describe how the susceptibility of local compression structures departs from this solvable limit.

\subsection{Weighted Curvature Measures and Finite-Dimensional Geometry}
\label{subsec:weighted_curvature_geometry}

So far, the curvature-conditioned point process has been formulated using the logarithmic curvature tensor associated with the entropy functional
\begin{align}
	\mathcal E(\rho)
	=
	\int
	\rho\ln\rho\,\pd^3x.
\end{align}
In this case, the induced geometric measure is defined with respect to the ordinary volume element $\pd^3x$.
However, the transport-geometric formulation naturally admits weighted measures that selectively amplify specific regions of the density field.
This extension leads to a finite-dimensional curvature geometry associated with concentration-sensitive statistics.

Let
\begin{align}
	\pd\mu_N
	=
	\rho(\bm{x})^N\,\pd^3x
	\label{eq:weighted_measure}
\end{align}
be a weighted measure indexed by the parameter $N$.
The logarithm of the corresponding weight is
\begin{align}
	\ln\rho^N
	=
	N\ln\rho,
\end{align}
and therefore the associated curvature tensor becomes
\begin{align}
	H_{ij}^{(N)}
	=
	-\partial_i\partial_j\ln\rho^N
	=
	NH_{ij}.
	\label{eq:weighted_curvature_tensor}
\end{align}
Thus, the finite-dimensional extension does not introduce a new geometric object independent of the logarithmic curvature tensor.
Instead, it rescales the sensitivity of the geometry to density concentration.

\begin{figure}[t]
	\centering
	\includegraphics[width=0.7\textwidth]{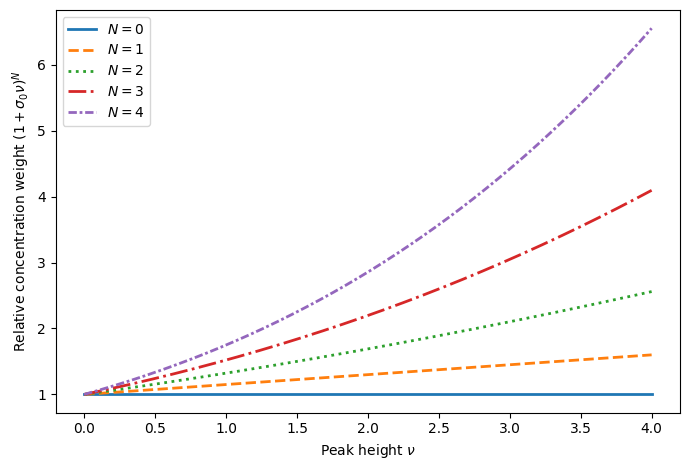}
    \includegraphics[width=0.7\textwidth]{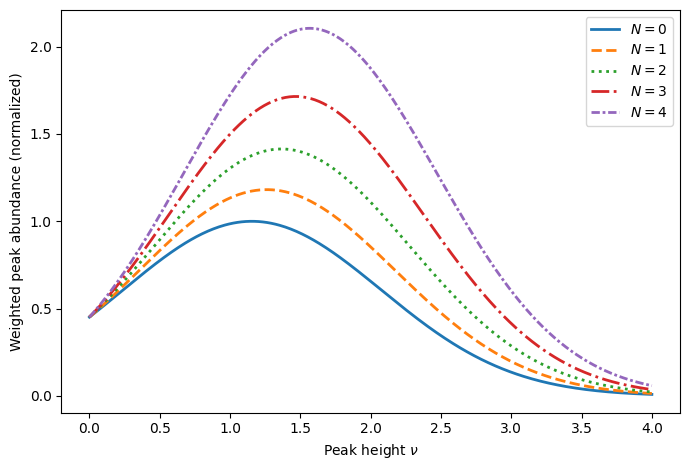}
	\caption{
	Illustration of concentration-sensitive weighting in the curvature-conditioned response hierarchy.
	The plotted curves show representative weighted peak abundances generated from the weighted measure defined in Eq.~\eqref{eq:weighted_peak_measure} for several values of the weighting parameter $N$.
	Larger values of $N$ progressively enhance the contribution of high-density transport structures relative to the unweighted Gaussian reference ensemble.
	The figure illustrates how concentration weighting deforms the geometric susceptibility of the response hierarchy while preserving the underlying curvature-conditioned point-process structure.
	}
	\label{fig:weighted_response}
\end{figure}

The corresponding weighted peak statistic is
\begin{align}
	n_{\rm pk}^{(N)}
	=
	\int
	\pd\mu_N\,
	|\det H|\,
	\deltadir^{(3)}(g)\,
	\Theta(H).
	\label{eq:weighted_peak_measure}
\end{align}
Substituting Eq.~\eqref{eq:weighted_measure},
\begin{align}
	n_{\rm pk}^{(N)}
	=
	\int
	\rho(\bm{x})^N
	\,\pd^3x\,
	|\det H|\,
	\deltadir^{(3)}(g)\,
	\Theta(H).
\end{align}
This expression shows explicitly that the weighted theory amplifies the contribution of high-density regions while suppressing low-density regions.
To illustrate the effect of concentration-sensitive weighting, we consider representative weighted peak statistics generated by Eq.~\eqref{eq:weighted_peak_measure}.
Figure~\ref{fig:weighted_response} shows the dependence of the weighted peak abundance on the parameter $N$ for a representative Gaussian reference spectrum.
As $N$ increases, the weighting progressively enhances high-density transport structures and amplifies the contribution of concentrated peaks relative to the unweighted BBKS reference ensemble.

The response structure is modified accordingly.
Differentiating Eq.~\eqref{eq:weighted_peak_measure} with respect to the background field gives
\begin{align}
	\partial_{\delta_L}
	n_{\rm pk}^{(N)}
	=
	I_{\rm weight}^{(N)}
	+
	I_{\rm det}^{(N)}
	+
	I_{\Theta}^{(N)},
\end{align}
where the new weighted contribution is
\begin{align}
	I_{\rm weight}^{(N)}
	=
	N
	\int
	\rho^{N-1}
	(\partial_{\delta_L}\rho)\,
	|\det H|\,
	\deltadir^{(3)}(g)\,
	\Theta(H)\,
	\pd^3x.
	\label{eq:weighted_response_term}
\end{align}
Thus, the weighted theory introduces an additional response channel associated directly with density concentration.

This modification has a clear geometric meaning.
In the ordinary formulation, all peaks contribute according to the same underlying volume measure.
In the weighted formulation, however, peaks embedded in highly concentrated regions acquire enhanced geometric weight.
The parameter $N$ therefore controls the degree to which the response hierarchy is biased toward concentrated transport structures.

The effect becomes particularly significant in nonlinear environments such as halo cores, cluster interiors, and strongly compressed filamentary regions.
In such environments, the logarithmic curvature tensor is already enhanced by nonlinear transport.
The weighted measure amplifies this enhancement further through the factor $\rho^N$.
Consequently, the weighted hierarchy selectively probes regions where local compression geometry and density concentration become strongly coupled.

The weighted formulation also modifies the effective response kernels.
For example, the one-point response becomes
\begin{align}
	R_{{\rm pk},N}^{(1)}
	=
	R_{\rm pk}^{(1)}
	+
	N
	\left\langle
		\frac{
			\partial_{\delta_L}\rho
		}{
			\rho
		}
	\right\rangle_{{\rm pk},N},
\end{align}
showing explicitly that concentration weighting generates an additional susceptibility proportional to the logarithmic density response.
Similarly, higher-order responses acquire weighted joint-curvature kernels emphasizing correlated high-density structures.

Importantly, this extension still preserves the curvature-conditioned skeleton of the theory.
The stationary-point condition, the cone constraint, and the determinant Jacobian remain unchanged.
What is modified is the geometric weighting assigned to different regions of the transport flow.
In this sense, the weighted theory should be understood not as a replacement of the curvature-conditioned point process, but as a concentration-sensitive deformation of its response geometry.

The finite-dimensional weighted measure therefore provides a natural extension of peak statistics from a theory of extrema to a theory of concentration-weighted geometric susceptibilities.
It introduces a continuous parameter controlling how strongly the response hierarchy emphasizes dense local compression structures generated by nonlinear transport.

\subsection{Concentration-Sensitive Peak Statistics}
\label{subsec:concentration_sensitive_statistics}

The weighted curvature formulation developed above changes the interpretation of peak statistics in an essential way.
In the ordinary Gaussian theory, peaks are primarily classified by amplitude and local curvature.
In the weighted theory, however, the statistical importance of a peak additionally depends on how strongly the surrounding mass distribution is concentrated.
As a result, the theory is naturally extended from a statistics of extrema to a statistics of concentration-sensitive geometric structures.

This distinction becomes particularly important in nonlinear structure formation.
In weakly nonlinear regions, the logarithmic curvature tensor differs only mildly from the linear Hessian.
By contrast, in strongly concentrated regions such as halo cores or collapsing filament intersections, the nonlinear gradient contributions become large and the density weighting factor $\rho^N$ selectively amplifies these structures.
Consequently, the weighted response hierarchy preferentially probes regions where transport-induced compression is strongest.

This effect can be seen explicitly by combining the determinant deformation derived in Section~\ref{sec:nonlinear_deformation} with the weighted measure.
Using
\begin{align}
	\det H
	=
	\det H^{(0)}
	\left[
		1+
		\sum_{a=1}^3
		\frac{
			(e_a\cdot\nabla\delta)^2
		}{
			\lambda_a^{(0)}
		}
		-
		3\delta
	\right]
	+
	O\!\left((H^{(2)})^2\right),
\end{align}
the weighted peak measure becomes
\begin{align}
	n_{\rm pk}^{(N)}
	=
	\int
	\rho^N
	|\det H^{(0)}|
	\left[
		1+
		\sum_{a=1}^3
		\frac{
			(e_a\cdot\nabla\delta)^2
		}{
			\lambda_a^{(0)}
		}
		-
		3\delta
	\right]
	\deltadir^{(3)}(g)\,
	\Theta(H)\,
	\pd^3x.
	\label{eq:weighted_nonlinear_peak_measure}
\end{align}
Equation~\eqref{eq:weighted_nonlinear_peak_measure} shows that concentration weighting and nonlinear curvature deformation reinforce one another.
The weighted hierarchy therefore amplifies precisely those structures for which nonlinear transport generates strong local compression geometry.

This interpretation is particularly useful from the viewpoint of transport geometry.
The logarithmic curvature tensor measures the local variation of the transport Jacobian, while the weighting factor $\rho^N$ amplifies regions where transported mass accumulates efficiently.
The weighted response hierarchy therefore probes not merely where peaks occur, but where transport-driven concentration is geometrically most effective.

The observational significance of this extension is also clear.
In practical galaxy and halo catalogs, the statistical importance of highly concentrated regions is often enhanced implicitly through selection effects, luminosity weighting, or halo-mass weighting.
The weighted curvature framework provides a geometric interpretation of such effects in terms of response weighting within the curvature-conditioned point process.
Thus, the parameter $N$ may be interpreted not merely as a formal mathematical extension, but as a phenomenological control parameter describing how strongly concentrated transport structures contribute to the observable response hierarchy.

The weighted theory also clarifies the relation between local geometry and large-scale environment.
Since the response coefficients depend on the background deformation $\delta_L$, the concentration weighting modifies how strongly dense local compression structures couple to long-wavelength modes.
As a result, the weighted hierarchy naturally generates environment-dependent geometric susceptibilities.
This provides a transport-geometric interpretation of why strongly concentrated structures are expected to exhibit enhanced large-scale response.

Taken together, the extensions developed in this section show that the curvature-conditioned point process admits a systematic deformation theory beyond the Gaussian-linear approximation.
Non-Gaussianity deforms the joint distribution of local tensor variables, while weighted measures deform the geometric sensitivity of the response hierarchy itself.
Importantly, however, neither extension alters the underlying local curvature-selection structure.
The zero-gradient condition, the positive-curvature constraint, and the determinant Jacobian remain the invariant geometric skeleton throughout the theory.

From this viewpoint, the classical BBKS formulation corresponds to a particularly simple sector of a much broader transport-geometric hierarchy.
The Gaussian-linear fixed-scale theory provides the analytically closed reference structure, while nonlinear transport, scale flow, non-Gaussian cumulants, and concentration weighting systematically deform the associated geometric susceptibilities.
The present framework therefore reorganizes peak statistics as a deformation theory of curvature-selected transport structures rather than as a collection of isolated extremum-counting formulas.

\subsection{Boundary Response of the Positive Curvature Cone}
\label{subsec:cone_boundary_response}

The previous sections showed that the response hierarchy of the curvature-conditioned point process decomposes into three sectors:
the distributional response, the determinant-measure response, and the selection-boundary response.
Among these, the boundary sector is particularly important because it governs how local configurations enter or leave the positive curvature cone under deformation.

The peak condition is defined by
\begin{align}
	H\in\mathcal C_+,
\end{align}
where $\mathcal C_+$ denotes the positive curvature cone in the symmetric matrix space $\mathrm{Sym}(3)$.
Equivalently, in eigenvalue space,
\begin{align}
	\lambda_a>0,
	\qquad
	a=1,2,3.
\end{align}
The boundary of the cone is therefore determined by the conditions
\begin{align}
	\lambda_a=0,
\end{align}
corresponding to configurations where at least one principal curvature vanishes.
Near this boundary, the classification of local structures becomes unstable under perturbation because infinitesimal deformations may move the curvature tensor into or out of the admissible region.

This structure appears directly in the response hierarchy through the derivative of the cone-selection factor,
\begin{align}
	\partial_{\delta_L}\Theta(H).
\end{align}
Writing the cone constraint in terms of eigenvalues,
\begin{align}
	\Theta(H)
	=
	\prod_{a=1}^3
	\Theta(\lambda_a),
\end{align}
its response becomes
\begin{align}
	\partial_{\delta_L}\Theta(H)
	=
	\sum_{a=1}^3
	\delta_{\rm D}(\lambda_a)
	\left(
		\partial_{\delta_L}\lambda_a
	\right)
	\prod_{b\neq a}
	\Theta(\lambda_b).
	\label{eq:cone_boundary_response}
\end{align}
Equation~\eqref{eq:cone_boundary_response} is one of the central consequences of the present framework.
It shows that the response of the selection geometry is localized on the boundary of the positive curvature cone.
Only configurations with nearly vanishing principal curvature contribute strongly to the boundary response.

This result has a clear geometric interpretation.
The determinant sector measures how the local volume element changes under deformation, while the cone-boundary sector measures how the admissible region itself changes.
Therefore, the response hierarchy contains not only a deformation of the geometric measure, but also a deformation of the geometric classification rule.

The effect becomes especially important near weakly curved structures.
Suppose that one eigenvalue satisfies
\begin{align}
	\lambda_a\simeq 0.
\end{align}
Then even a small background perturbation can change the sign of $\lambda_a$ and thereby alter whether the configuration is classified as a peak.
Consequently, the response of the peak abundance receives enhanced contributions from nearly degenerate curvature configurations.

This sensitivity is amplified further by the determinant response.
From Eq.~\eqref{eq:det_response_eigen},
\begin{align}
	\partial_{\delta_L}\ln|\det H|
	\simeq
	\sum_{a=1}^3
	\frac{
		e_a^i
		(\partial_{\delta_L}H_{ij})
		e_a^j
	}{
		\lambda_a
	},
\end{align}
so the determinant sector itself becomes large near the cone boundary.
Therefore, weakly curved configurations are simultaneously amplified by the determinant response and by the cone-boundary response.

The geometric significance of this result is important.
In the Gaussian fixed-scale theory, the peak condition is treated as a static constraint.
In the present formulation, however, the admissible region in curvature tensor space itself responds dynamically to background deformation, nonlinear transport, and scale flow.
Thus, the classification of peaks is not fixed once and for all, but evolves under transport-induced curvature deformation.

This point also clarifies the relation between the curvature flow developed in Section~\ref{sec:peaks} and the response hierarchy developed in Section~\ref{sec:response_hierarchy}.
The scale flow transports curvature tensors through $\mathrm{Sym}(3)$, while the cone-boundary response determines how this transport modifies the admissible peak region.
Peak creation, disappearance, and reclassification under coarse graining therefore correspond geometrically to boundary-crossing events in curvature space.

The same mechanism extends naturally to higher-order statistics.
For example, in the two-point response function, correlated boundary crossings occur when pairs of local curvature tensors simultaneously approach the cone boundary.
The corresponding response kernels contain terms of the form
\begin{align}
	\delta_{\rm D}(\lambda_a^{(1)})
	\delta_{\rm D}(\lambda_b^{(2)}),
\end{align}
which probe correlated geometric instabilities of local compression structures.
Similarly, higher-order responses probe collective boundary sensitivity of increasingly complex curvature configurations.

From this viewpoint, the positive curvature cone is not merely a technical device selecting maxima.
It becomes a dynamical geometric object whose boundary controls the susceptibility of the curvature-conditioned point process.
The response hierarchy therefore contains, in addition to distributional and measure deformations, a genuine dynamics of geometric selection itself.

This structure goes beyond the standard Kac--Rice interpretation of extrema statistics.
The determinant Jacobian and the zero-gradient condition remain present, but the present theory additionally identifies the response and flow of the admissible curvature domain as an independent geometric sector.
In this sense, the transport-geometric formulation promotes the peak-selection condition from a static constraint to a dynamical component of the response hierarchy.

\subsection{Cone-Boundary Effects and the Breakdown of Perturbative Curvature Transport}\label{subsec:cone_boundary_effects}

The residual deviation visible in Figure~\ref{fig:grf_determinant_deformation} at large peak height $\nu$ suggests that the nonlinear deformation of the curvature-conditioned point process is not exhausted by the leading determinant correction alone.
In the present formulation, two qualitatively different effects are expected to contribute in the high-$\nu$ regime.

The first effect originates from higher-order nonlinear transport terms in the logarithmic curvature expansion.
Expanding
\begin{align}
	\ln(1+\delta)
	=
	\delta
	-
	\frac12\delta^2
	+
	\frac13\delta^3
	+\cdots,
\end{align}
the logarithmic curvature tensor contains cubic-order contributions such as $	\delta^2\partial_i\partial_j\delta$, $\delta	(\partial_i\delta)	(\partial_j\delta)$, in addition to the quadratic corrections discussed in Section~\ref{subsec:nonlinear_correction_1st}.
These terms generate further deformation of the determinant transport, modify the local curvature anisotropy, and induce higher-order shifts of the curvature-conditioned measure.
From this viewpoint, the deviation at high $\nu$ partly reflects the breakdown of the leading perturbative transport approximation.

However, the geometric structure developed in this work suggests that the discrepancy may not be explained solely by higher-order perturbative corrections.
In the high-$\nu$ regime, the smallest eigenvalue of the curvature tensor may approach the boundary of the positive-curvature cone,
\begin{align}
	\lambda_{\min}\rightarrow 0.
\end{align}
Near this boundary, the curvature-conditioned measure becomes increasingly sensitive to the geometry of $\mathcal{C}_+$ itself.
Consequently, the nonlinear deformation may acquire genuinely nonperturbative contributions associated with boundary effects of the curvature cone.

This distinction is conceptually important.
The first contribution corresponds to perturbative nonlinear transport inside curvature space, while the second reflects deformation of the admissible geometric domain itself.
In this sense, the high-$\nu$ discrepancy may provide a direct indication of where the perturbative expansion of the curvature-conditioned measure begins to fail geometrically.

The present framework therefore suggests a natural transition between perturbative curvature transport and nonperturbative cone-boundary geometry.
This interpretation is particularly interesting because the positive-curvature cone introduced in Section~2 is not merely a technical selection condition, but may control the onset of genuinely nonlinear geometric effects beyond Gaussian closure.

\subsection{Primordial Non-Gaussianity as Deformation of the Curvature-Conditioned Measure}
\label{subsec:png_curvature_measure}

The response hierarchy developed above naturally suggests a geometric interpretation of primordial non-Gaussianity.
In conventional bias theory, local-type primordial non-Gaussianity is usually introduced as an external modulation of the long-wavelength potential,
\begin{align}
	\Phi
	=
	\phi_G
	+
	f_{\rm NL}
	\left(
	\phi_G^2
	-
	\langle
	\phi_G^2
	\rangle
	\right),
\end{align}
which induces scale-dependent corrections to halo or peak bias \citep[e.g.,][]{2008PhRvD..77l3514D}.
In such approaches, the effect of $f_{\rm NL}$ appears primarily as an additional response of tracers to a background potential field.

In the present formulation, however, the geometric interpretation is different.
Since the fundamental object is not a bias function but a curvature-conditioned probability measure on local tensor space, primordial non-Gaussianity can instead be regarded as a deformation of the curvature-space measure itself,
\begin{align}
	p(H)
	\rightarrow
	p(H)
	+
	f_{\rm NL}\Delta p(H).
\end{align}
From this viewpoint, the effect of primordial non-Gaussianity is not merely to modulate the abundance of peaks externally.
Rather, it changes the intrinsic probability distribution of curvature configurations inside the positive-curvature cone $\mathcal{C}_+$.

This distinction is geometrically important.
The conventional picture resembles an external forcing acting on an otherwise fixed fluid configuration.
By contrast, in the transport-geometric formulation developed here, the deformation originates from the initial probability measure itself.
That is, the primordial state of the density field is interpreted as having an intrinsically distorted curvature geometry in Wasserstein space.
Consequently, the response hierarchy is naturally extended from
\begin{align}
	R_{\rm pk}^{(n)}
	=
	\frac{
	\delta^n n_{\rm pk}
	}{
	\delta \delta_L^n
	}
\end{align}
to a hierarchy of deformations of the curvature-conditioned measure itself.

This viewpoint also suggests that primordial non-Gaussianity may induce deformations not only of the peak abundance, but also of the eigenvalue structure and boundary geometry of the positive-curvature cone.
In particular, correlations such as $\langle HHH \rangle$, $	\langle	\nu x x \rangle$, which vanish in the Gaussian closure, become nonzero and generate anisotropic deformations of the local curvature distribution.
Thus, primordial non-Gaussianity appears not simply as a correction to bias coefficients, but as a geometric deformation of the curvature-conditioned point process itself.

From this perspective, the response hierarchy derived in this section may provide a natural framework connecting optimal transport geometry, peak statistics, and primordial non-Gaussianity within a unified measure-theoretic description.

\section{Discussion and Outlook}
\label{sec:discussion}

In this work, we have reformulated the peak statistics of cosmological density fields within the framework of curvature statistics based on optimal transport and entropy geometry.
The starting point is to regard the density field not as a function but as a probability measure, and to describe its deformation as a transport in Wasserstein geometry.
In this formulation, the Hessian of the logarithmic density naturally appears as the local response of the entropy functional,
\begin{align}
	H_{ij}=-\partial_i\partial_j\ln\rho \notag
\end{align}
which serves as the fundamental definition of the curvature tensor in this paper.

Based on this curvature tensor, peaks are no longer merely density maxima, but are geometric objects defined as points satisfying
\begin{align}
	g_i=0,\qquad H\in\mathcal{C}_+ \notag
\end{align}
where $g_i=\partial_i\ln\rho$ is the logarithmic density gradient introduced in Section~\ref{sec:curvature_geometry}, which reduces to $\eta_i=\partial_i\delta$ used in Section~\ref{sec:peaks} in the linear Gaussian limit.
Furthermore, the peak number density is given as a curvature-conditioned point process,
\begin{align}
	n_{\rm pk}
	=
	\int \pd\rho\,\pd^3 g\,\pd H\,
	|\det H|\,
	p(\rho,g,H)\,
	\deltadir^{(3)}(g)\,
	\Theta(H) \notag
\end{align}
As this expression shows, the essence of peak statistics lies in the combination of three elements,
\begin{align}
    \begin{array}{c}
        \text{probability distribution}\\[4pt]
        \times\\[4pt]
        \text{curvature constraint}\\[4pt]
        \times\\[4pt]
        \text{geometric measure}
    \end{array}
    \notag
\end{align}

This general structure does not depend on Gaussianity.
The assumption of a Gaussian random field only serves to close the joint distribution $p(\rho,g,H)$ in terms of a finite number of spectral moments.
Therefore, the BBKS theory is positioned as a special case in which explicit evaluation becomes possible under the linear Gaussian condition, within the general theory of curvature-conditioned point processes.

In this paper, we have further extended this framework to $n$-point statistics.
The two-point and three-point peak correlations are defined as
\begin{align}
	\rho_{\rm pk}^{(2)}(\bm{x}_1,\bm{x}_2)
	&=
	\left\langle
		n_{\rm pk}(\bm{x}_1)n_{\rm pk}(\bm{x}_2)
	\right\rangle, \\	
	\rho_{\rm pk}^{(3)}(\bm{x}_1,\bm{x}_2,\bm{x}_3)
	&=
	\left\langle
		\prod_{a=1}^3 n_{\rm pk}(\bm{x}_a)
	\right\rangle
\end{align}
and are expressed as conditional integrals over the joint distribution of the curvature tensor field.
Through this structure, peak statistics are understood not as mere positional statistics, but as geometric statistics describing the spatial correlations of local compression structures.

The final outcome of this work is that these $n$-point statistics can be understood in a unified manner as a hierarchy of response functions.
Defining the response to a long-wavelength background field $\delta_{\rm L}$ as
\begin{align}
	R_{\rm pk}^{(n)}
	=
	\pa{\ln \rho_{\rm pk}^{(n)}}{\delta_{\rm L}},
\end{align}
one obtains the response hierarchy
\begin{align}
	R_{\rm pk}^{(1)}=b_{\rm pk},\quad
	R_{\rm pk}^{(2)},\quad
	R_{\rm pk}^{(3)},\ \dots
\end{align}
Thus, the entire peak theory is reconstructed as a response-function theory of curvature-selected point processes.

Within this reconstruction, the position of the BBKS theory becomes clear.
Namely, BBKS corresponds to a closed-form solution obtained under the conditions
\begin{align}
    \begin{array}{c}
        \text{curvature-conditioned point process}\\[4pt]
        +\\[4pt]
        \text{linear limit}\\[4pt]
        +\\[4pt]
        \text{Gaussian closure}\\[4pt]
        +\\[4pt]
        \text{fixed scale}
    \end{array}
\end{align}
It therefore represents one slice of the general theory developed in this paper.
Accordingly, the present framework naturally allows extensions to general curvature statistics beyond the linear Gaussian regime:
\begin{enumerate}
	\item By defining peaks in terms of logarithmic density curvature, nonlinear structures are described in the same formalism.
	\item Non-Gaussianity is incorporated as a deformation of the joint distribution $p(\rho,g,H)$, while the three-element decomposition remains intact.
	\item By introducing a smoothing scale $R$, curvature statistics can be understood as flows in scale space.
\end{enumerate}
Furthermore, by introducing the weighted measure
\begin{align}
	\pd\mu_N=\rho(\bm{x})^N\pd^3 x \notag
\end{align}
new statistical quantities based on finite-dimensional curvature can be defined.

As future directions, it is important to derive explicit evaluation formulas for non-Gaussian distributions.
In particular, it is essential to clarify how primordial non-Gaussianity and nonlinear gravitational evolution appear in the joint distribution of the curvature tensor and in the response-function hierarchy.
In this context, comparison with recent studies that provide systematic perturbative expansions of statistical quantities \citep[e.g.,][]{2024PhRvD.110f3543M} is of interest.
While the framework of this paper is a geometric formulation that does not rely on perturbative expansion, the two approaches are complementary from the viewpoint of explicit expansion of joint distributions and evaluation of response functions, and their integration is expected to lead to further theoretical developments.
Moreover, by constructing explicit equations for scale flow, it becomes possible to describe structure formation as a dynamical system in curvature geometry.
Observationally, measurements of curvature statistics for galaxy and halo distributions constitute an important application.
In particular, formulations that incorporate finite-sample effects and observational window functions are intrinsically consistent with the measure-theoretic framework developed in this paper.

In conclusion, the present work reconstructs peak theory from a mere statistics of extrema into a geometric response theory based on optimal transport and entropic curvature.
This unified perspective interprets cosmological structure formation as the transport of probability measures and positions peak statistics as the response of local compression structures.
This framework connects statistical and geometric methods, and is expected to open new directions for both theoretical and observational studies.

\section*{Acknowledgments}

I am deeply grateful to Shiro Ikeda, Satoshi Kuriki, and Dan Cheng for helpful insights for this work.
This work was supported by JSPS Grant-in-Aid for Scientific Research (24H00247) and by the Joint Research Program of the Institute of Statistical Mathematics (General Research 2), ``Machine-learning cosmogony: from structure formation to galaxy evolution.''

\bibliographystyle{JHEP}
\bibliography{BBKS_Wasserstein}

\appendix

\section{Post-Publication Clarification:
Exact Cancellation in the Complete Kac--Rice Peak Operator}

Sections~\ref{subsec:determinant_peak_measure}
and~\ref{subsec:nonlinear_peak_abundance} of the published version isolate the deformation of the Hessian-determinant factor produced by replacing the Gaussian-linear curvature tensor with the logarithmic curvature tensor.  That calculation remains valid as a determinant-sector response.  A subsequent examination of the complete
Kac--Rice operator shows, however, that this partial response does not by itself imply a change in the unmarked peak abundance.

Let
\begin{align}
\rho
&=
\bar{\rho}(1+\delta)
>
0,
\qquad
\phi
=
\ln\rho,
\notag\\
\bm{\eta}
&\equiv
\nabla\delta,
\qquad
H^{(0)}
\equiv
-\nabla\nabla\delta.
\end{align}
The logarithmic gradient and curvature tensor are then
\begin{align}
\bm{g}_{\phi}
&\equiv
\nabla\phi
=
\frac{\bm{\eta}}{1+\delta},
\label{eq:clarification_log_gradient}
\\
H_{\phi}
&\equiv
-\nabla\nabla\phi
=
\frac{H^{(0)}}{1+\delta}
+
\frac{
\bm{\eta}\bm{\eta}^{\top}
}{
(1+\delta)^2
}.
\label{eq:clarification_log_hessian}
\end{align}
Because $1+\delta>0$, the stationary-point conditions are exactly
equivalent:
\begin{align}
\bm{g}_{\phi}
=
0
\quad\Longleftrightarrow\quad
\bm{\eta}
=
0.
\label{eq:clarification_stationary_equivalence}
\end{align}
On this stationary set, the rank-one gradient term in
Eq.~\eqref{eq:clarification_log_hessian} vanishes and
\begin{align}
H_{\phi}
&=
\frac{H^{(0)}}{1+\delta}.
\label{eq:clarification_stationary_hessian}
\end{align}
Consequently, in $d$ spatial dimensions,
\begin{align}
\left|
\det H_{\phi}
\right|
&=
(1+\delta)^{-d}
\left|
\det H^{(0)}
\right|,
\label{eq:clarification_determinant_scaling}
\\
\delta_{\mathrm D}^{(d)}
\left(
\bm{g}_{\phi}
\right)
&=
(1+\delta)^d
\delta_{\mathrm D}^{(d)}
\left(
\bm{\eta}
\right).
\label{eq:clarification_delta_scaling}
\end{align}
Moreover, multiplication of a symmetric matrix by the positive scalar
$(1+\delta)^{-1}$ preserves its inertia.  Hence, denoting by
$\Theta_{+}(H)$ the indicator of the positive-curvature cone,
\begin{align}
\Theta_{+}
\left(
H_{\phi}
\right)
&=
\Theta_{+}
\left(
H^{(0)}
\right)
\end{align}
on the stationary set.

The complete unweighted Kac--Rice peak operator therefore satisfies
the exact identity
\begin{align}
&
\left|
\det H_{\phi}
\right|
\delta_{\mathrm D}^{(d)}
\left(
\bm{g}_{\phi}
\right)
\Theta_{+}
\left(
H_{\phi}
\right)
\notag\\
&\qquad
=
\left|
\det H^{(0)}
\right|
\delta_{\mathrm D}^{(d)}
\left(
\bm{\eta}
\right)
\Theta_{+}
\left(
H^{(0)}
\right).
\label{eq:clarification_kac_rice_identity}
\end{align}
Thus, the determinant rescaling and the Jacobian of the
stationary-point delta function cancel exactly.  In perturbative form,
\begin{align}
(1+\delta)^{-d}
&=
1
-
d\delta
+
\frac{d(d+1)}{2}\delta^2
+
O(\delta^3),
\notag\\
(1+\delta)^d
&=
1
+
d\delta
+
\frac{d(d-1)}{2}\delta^2
+
O(\delta^3),
\end{align}
and hence
\begin{align}
&
\left[
1
-
d\delta
+
\frac{d(d+1)}{2}\delta^2
+
\cdots
\right]
\notag\\
&\qquad\times
\left[
1
+
d\delta
+
\frac{d(d-1)}{2}\delta^2
+
\cdots
\right]
=
1.
\label{eq:clarification_order_by_order_cancellation}
\end{align}
The cancellation therefore holds order by order in any consistent perturbative expansion of the complete Kac--Rice operator.  In particular, for $d=3$, the leading determinant-sector factor $-3\delta$ is canceled by the corresponding $+3\delta$ contribution from the stationary-point delta-function Jacobian.  Terms containing $\bm{\eta}$ also vanish on the support of $\delta_{\mathrm D}^{(d)}(\bm{\eta})$.

Accordingly, the determinant-sector corrections displayed in the published article should be interpreted as partial deformations of the curvature-mark measure, rather than as net corrections to the complete unmarked Kac--Rice peak abundance.  The associated determinant deformation remains a valid diagnostic of the log-curvature mark.  At each peak, that mark is transformed exactly as
\begin{align}
H^{(0)}
&\longmapsto
H_{\phi}
=
\frac{H^{(0)}}{1+\delta}.
\label{eq:clarification_curvature_mark_transport}
\end{align}
The peak locations, their Morse classification, and the unmarked peak count remain unchanged by the smooth positive pointwise reparameterization $\rho\mapsto\ln\rho$ at fixed smoothing scale.

This clarification does not affect the exact recovery of the Gaussian BBKS formula or the curvature-conditioned marked-point framework.
A genuine change in the unmarked peak abundance requires an operation that changes the underlying field or its selection rule, such as scale-dependent smoothing, a nonlocal finite-window construction, or an explicitly modified peak weighting, rather than a smooth monotone
pointwise reparameterization alone.

\section{Eigenvalue Representation of the Curvature Measure}
\label{app:eigenvalue_measure}

In the main text, the Gaussian closure of the curvature-conditioned
point process was shown to lead to the BBKS peak abundance formula.
The purpose of this Appendix is to outline the geometric reduction
of the curvature measure that underlies this result.

\subsection{Measure on the Space of Curvature Tensors}

The logarithmic curvature tensor$H_{ij}$ is a real symmetric $3\times3$ matrix and therefore belongs to $\mathrm{Sym}(3)$.
The corresponding integration measure may be written as
\begin{align}
	\pd H
	=
	\prod_{i\le j}
	\pd H_{ij}.
\end{align}
For the purpose of evaluating peak statistics, it is convenient
to diagonalize the curvature tensor,
\begin{align}
	H
	=
	O^\top
	\Lambda
	O,
\end{align}
where $O\in{\rm O}(3)$ is an orthogonal matrix and
\begin{align}
	\Lambda
	=
	{\rm diag}
	(\lambda_1,\lambda_2,\lambda_3)
\end{align}
contains the eigenvalues of $H$.

Under this transformation, the measure becomes
\begin{align}
	\pd H
	=
	|\Delta(\lambda)|
	\,
	\pd^3\lambda
	\,
	\pd\mu(O),
	\label{eq:eigen_measure}
\end{align}
where
\begin{align}
	\pd^3\lambda
	=
	\pd\lambda_1
	\pd\lambda_2
	\pd\lambda_3,
\end{align}
$\pd\mu(O)$ denotes the invariant Haar measure on the orthogonal
group, and
\begin{align}
	\Delta(\lambda)
	=
	(\lambda_1-\lambda_2)
	(\lambda_1-\lambda_3)
	(\lambda_2-\lambda_3)
\end{align}
is the Vandermonde determinant.

\subsection{Geometry of Eigenvalue Space}

Equation~(\ref{eq:eigen_measure}) separates the rotational degrees
of freedom from the intrinsic shape of the curvature tensor.
The orthogonal matrix $O$ specifies the local orientation of the
principal axes, while the eigenvalues $\lambda_i$ characterize the
local curvature configuration itself.
The Vandermonde factor $|\Delta(\lambda)|$ has an important geometric interpretation.
Configurations with degenerate eigenvalues occupy a lower-dimensional
subset of $\mathrm{Sym}(3)$ and therefore carry vanishing measure.
Consequently, the induced measure naturally suppresses curvature
configurations with coincident principal curvatures.
From this viewpoint, peak statistics are not merely determined by
curvature amplitudes, but also by the geometry of eigenvalue space.

\subsection{Positive-Curvature Cone}

The peak condition requires the logarithmic curvature tensor to be
positive definite.
In eigenvalue space, this condition becomes
\begin{align}
	\lambda_i>0,
	\qquad
	i=1,2,3.
\end{align}
The allowed domain therefore corresponds to the
positive-curvature cone
\begin{align}
	\mathcal C_+
	=
	\{
	H\in\mathrm{Sym}(3)
	\mid
	\lambda_i>0
	\}.
\end{align}
This restriction is independent of Gaussianity and constitutes
one of the fundamental geometric ingredients of the
curvature-conditioned point process.

\subsection{Reduction of the Peak Measure}

The Gaussian peak abundance is obtained from
\begin{align}
	n_{\rm pk}(\nu)
	=
	\int
	\pd H
	|\det H|
	p(\nu,H\,|\,\eta=0)
	\Theta(H).
\end{align}
Using Eq.~(\ref{eq:eigen_measure}) and integrating over the
rotational degrees of freedom, one obtains
\begin{align}
	n_{\rm pk}(\nu)
	\propto
	\int_{\lambda_i>0}
	\pd^3\lambda
	|\lambda_1\lambda_2\lambda_3|
	|\Delta(\lambda)|
	p(\nu,\lambda_1,\lambda_2,\lambda_3).
\end{align}
The peak abundance is therefore determined by three geometric
ingredients:
\begin{itemize}
    \item curvature amplitude, 
    \item eigenvalue configuration, 
    \item positive-curvature selection.
\end{itemize}
After expressing the distribution in terms of the standard BBKS
variables $(\nu,x)$ and carrying out the remaining integrations,
one recovers Eq.~(\ref{eq:BBKS_recovered}) in the main text.
The remaining integrations depend only on the two
dimensionless combinations
\begin{align}
	\nu=\frac{\delta}{\sigma_0},
	\qquad
	x=-\frac{\nabla^2\delta}{\sigma_2},
\end{align}
and can therefore be reduced to the standard BBKS form.
After expressing the distribution in terms of these variables and
carrying out the remaining integrations, one recovers
Eq.~(\ref{eq:BBKS_recovered}) in the main text.

\end{document}